\documentclass[twocolumn]{aastex6}
\usepackage{subfigure}
\newcommand{\tctff}{{t$_{\rm cool}$/t$_{\rm ff}$}}
\newcommand{\tctffmin}{(\tctff)$_{\rm min}$}
\newcommand{\Rtctffmin}{R(\tctff)$_{\rm min}$}

\newcommand{\tc}{$t_{\rm cool}$}
\newcommand{\tccentral}{$t_{\rm c, central}$}
\newcommand{\tff}{$t_{\rm ff}$}

\fullcollaborationName{The Friends of AASTeX Collaboration}

\begin{document}

\title{The Onset of Thermally Unstable Cooling from the Hot Atmospheres of Giant Galaxies in Clusters -- Constraints on Feedback Models}

\author{M.~T. Hogan\altaffilmark{1,2}, B.~R. McNamara\altaffilmark{1,2}, F.~A. Pulido\altaffilmark{1}, P.~E.~J. Nulsen\altaffilmark{3,4}, A.~N. Vantyghem\altaffilmark{1}, H.~R. Russell\altaffilmark{5}, \\ A.~C. Edge\altaffilmark{6}, Iu. Babyk\altaffilmark{1,7}, R.~A. Main\altaffilmark{8}, and M. McDonald\altaffilmark{9}}

\altaffiltext{1}{Department of Physics and Astronomy, University of Waterloo, Waterloo, ON, N2L 3G1, Canada}
\altaffiltext{2}{Perimeter Institute for Theoretical Physics, Waterloo, ON, N2L 2Y5, Canada}
\altaffiltext{3}{Harvard-Smithsonian Center for Astrophysics, 60 Garden Street, Cambridge, MA 02138, USA}
\altaffiltext{4}{ICRAR, University of Western Australia, 35 Stirling Hwy, Crawley, WA 6009, Australia}
\altaffiltext{5}{Institute of Astronomy, Madingley Road, Cambridge, CB3 0HA, UK}
\altaffiltext{6}{Centre for Extragalactic Astronomy, Department of Physics, Durham University, Durham, DH1 3LE, UK}
\altaffiltext{7}{Main Astronomical Observatory of NAS of Ukraine, 27 Academica Zabolotnogo Str, 03143, Kyiv, Ukraine}
\altaffiltext{8}{Canadian Institute for Theoretical Astrophysics, University of Toronto, 60 St. George Street, Toronto, ON, M5S 3H8, Canada}
\altaffiltext{9}{Kavli Institute for Astrophysics and Space Research, Massachusetts Institute of Technology, 77 Massachusetts Avenue, Cambridge, MA 02139, USA}

\email{m4hogan@uwaterloo.ca}

\begin{abstract}
We present accurate mass and thermodynamic profiles for a sample of 56 galaxy clusters observed with the {\it Chandra} X-ray Observatory. We investigate the effects of local gravitational acceleration in central cluster galaxies, and we explore the role of the local free-fall time (\tff) in thermally unstable cooling.  We find that the local cooling time (\tc) is as effective an indicator of cold gas, traced through its nebular emission, as the ratio of \tctff.   Therefore, \tc\ alone apparently governs the onset of thermally unstable cooling in hot atmospheres.  The location of the minimum \tctff, a thermodynamic parameter that simulations suggest may be key in driving thermal instability, is unresolved in most systems.  As a consequence, selection effects bias the value and reduce the observed range in measured \tctff\ minima. The entropy profiles of cool-core clusters are characterized by broken power-laws down to our resolution limit, with no indication of isentropic cores.  We show, for the first time, that mass isothermality and the $K \propto r^{2/3}$ entropy profile slope imply a floor in \tctff\ profiles within central galaxies.  No significant departures of \tctff\ below 10 are found, which is inconsistent with many recent feedback models.  The inner densities and cooling times of cluster atmospheres are resilient to change in response to powerful AGN activity, suggesting that the energy coupling between AGN heating and atmospheric gas is gentler than most models predict. 
\end{abstract}


\section{Introduction} \label{Section:Introduction}

The hot atmospheres at the centers of many galaxies and galaxy clusters radiate X-rays so prodigiously they are expected to cool on timescales much shorter than their ages.  Unless radiation losses are compensated by heating, their central atmospheres would cool at rates of hundreds to thousands of solar masses per year and form stars \cite[for a review see][]{Fabian94}.  Observations have instead shown far less molecular gas \cite[][]{Edge01,Salome03}, star formation  \cite[][]{Johnstone87,O'Dea08,Rafferty08}, and cooling gas \cite[][]{Peterson03,Borgani06,Nagai07,Sanders11} than expected.  Cooling must therefore be suppressed.  Observation has shown that mechanical feedback from the active galactic nucleus (AGN) within the centrally located brightest cluster galaxy (BCG) to be the most likely mechanism \cite[][]{McNamara07}.

In the standard picture of AGN Feedback, radio jets launched by supermassive black holes (SMBH) inflate cavities that rise buoyantly through the intracluster medium (ICM) driving turbulence, shocks, and sound waves \cite[][]{Fabian05,Voit05b,Randall11,Nulsen13,Zhuravleva14,Hillel16a,Hillel16b,Soker16,Yang16a}.  The enthalpy released by AGN raises the entropy of the surrounding atmosphere and regulates the rate of cooling \cite[for reviews see][]{McNamara07,McNamara12,Fabian12}.  

Cooling into molecular clouds must occur in order to maintain the feedback cycle.  Observations of molecular gas \cite[][]{Edge01,Salome03}, nebular emission \cite[e.g.][]{Heckman89,Crawford99,McDonald10,Tremblay15}, and star formation, are indeed observed at levels consistent with having been fueled by cooling from the surrounding hot atmosphere \cite[][]{McNamara14,Russell17}.   Feedback is apparently persistent.   Cool-core clusters have existed for at least half the age of the Universe \cite[e.g.][]{Santos10,Samuele11,Ma11,McDonald13,Hlavacek-Larrondo15,Main17}.  Their prevalance requires long-term equilibrium between heating and cooling \cite[e.g.][]{Hlavacek-Larrondo12, Main17}, despite large variations of power output from their AGN \cite[][]{Hogan15b}.

Nebular emission, increased star-formation, and AGN activity are preferentially observed in cluster cores when the central entropy $K$ drops below 30~keV~cm$^{2}$, roughly equivalent to a central cooling time less than 1~Gyr \cite[][]{Cavagnolo08,Rafferty08,Sanderson09a,Main17}.  Though this threshold is sharp, in our view a convincing physical understanding of cooling instability at the centers of giant galaxies remains elusive.  This threshold accurately presages molecular gas in central galaxies at levels far above those seen in normal ellipticals, (Pulido et al {\em in prep.}, henceforth Paper II), and this molecular gas is likely fueling the AGN feedback cycle \cite[][]{Tremblay16}.  While our understanding of AGN heating has matured, our understanding of thermally unstable cooling is less advanced, yet it is the other crucial aspect of the feedback cycle. 

\subsection{Review of \tctff\ Models and Observations}

Hot atmospheres are thought to become thermally unstable in their central regions when the ratio of the cooling time, \tc, to the free-fall time, \tff, of a parcel of cooling gas falls below unity \cite[][]{Cowie80,Nulsen86}.  Interest in this problem was revived recently by important papers showing that the instability criterion \tctff$\lesssim 1$ applies to gas cooling in a simulated plane-parallel atmosphere, but may rise well above unity in a three dimensional atmosphere \cite[][]{McCourt12,Sharma12b}.  These developments are potentially significant because the ratio \tctff\ never falls toward unity locally in central cluster galaxies, even when the atmosphere is cooling rapidly into molecular clouds and fueling star formation.  

Understanding how thermally unstable cooling is triggerd in clusters is essential because cold accretion likely plays a crucial role in the regulation of AGN feedback that may govern the growth of all massive galaxies \cite[][]{Gaspari12,Gaspari13,Li14b,Li15,Voit15a,Voit15b}.  However,  feedback involves complex physical interactions operating over many decades in scale, which is notoriously difficult to model.  Nevertheless, modern, high-fidelity, simulations have yielded predictions that can be tested using precision measurements, which is the focus of this paper. 

Three dimensional feedback simulations attempting to model the consequences of thermally unstable cooling that develops at a threshold of \tctff$\approx 10$ indicate that this mechanism may lead to a self-regulating feedback cycle \cite[][]{Gaspari12,Gaspari13,Prasad15,Li14a,Li15,Meece15,Singh15,Gaspari15}. These models, broadly referred to as ‘precipitation’ or ‘chaotic cold accretion,’ fuel the AGN and star formation that in turn suppresses further condensation leading to self regulation \cite[][]{Pizzolato05,Pizzolato10,Gaspari12,Li14a}.  

The mechanism works generally as follows:  thermally unstable cooling is assumed to occur when \tctff\ falls below $\sim$10.   The cooling gas then fuels both star formation and the AGN. As the radio AGN heats the atmosphere it lowers the central gas density, which in turn increases \tc\ in response to AGN heating \cite[e.g.][]{Li15,Voit15a}.  As the ratio \tctff\ rises above 10, thermally unstable condensation ceases, cutting off the fuel supply for the AGN and quenching feedback.  Over time, the atmosphere once again begins to cool.  

Repeated episodes of heating and cooling are thought to maintain \tctff\ above 10.  However, a key and testable aspect of these models is that the minimum value of \tctff\ in those systems experiencing a cooling episode should lie below 10.  Simulation shows that a significant fraction of the population at any given time should be in a minimum state below 10 \cite[e.g.][]{Li15}.  Furthermore, the ratio \tctff\ at its minimum value should predict the onset of thermally unstable cooling, as traced by nebular emission, molecular clouds, and star formation, with greater certainty and lower observational scatter than the local cooling time alone \cite[e.g.][]{Rafferty08}.  In other words, the additional physics associated with the denominator should act to decrease the scatter, if local acceleration is playing a significant role.  
This issue was addressed by \citet[][]{McNamara16} who showed that \tff\ at the location of the \tctff\ minimum spans only a narrow range of values in central galaxies.  They further showed that the ratio \tctff\ is driven almost entirely by \tc.  While these results taken at face value do not exclude a significant role for local acceleration, they imply that critical aspects of precipitation models are difficult to falsify, and thus may not be unique.  More significantly, they showed, as we do here,  that the inner gas densities of cooling atmospheres vary over a strikingly small range in response to an enormous  range of AGN power.  The muted response to AGN heating is both surprising and troubling for many AGN feedback models.

These considerations, in part, led \citet[][]{McNamara14,McNamara16} to suggest that thermally unstable cooling instead occurs when low entropy gas from the cluster center is lifted in the updraft of buoyantly-rising X-ray cavities.  Furthermore, ALMA observations have shown that molecular gas in cluster cores lies preferentially in the wakes of buoyantly rising cavities \cite[e.g.][]{McNamara14,Vantyghem16,Russell17}.  Whether the molecular gas is condensing directly from the uplifted hot gas, or whether the cold gas is being lifted directly is unclear.  However, indications are that at least some is cooling directly out of the hot atmosphere, at altitudes where the local value of \tctff\ greatly exceeds 10.   Moreover, numerical simulations have shown that marginally stable gas can be triggered to condense when uplifted by an AGN, indicating that this “stimulated feedback” mechanism is plausible at least \cite[][]{Revaz08,Gaspari12,Li14b,Brighenti15,Voit16,Yang16a}.

Despite the uncertain role of local acceleration, halo mass is clearly relevant to the AGN feedback cycle in clusters \cite[e.g][]{Main17}, and should be further explored.  In order to do so,  \citet[][]{Hogan17} developed techniques to determine cluster mass profiles across wide radial ranges that extend from cluster halos into the cores of the central galaxies.  We adopt this methodology in this paper to calculate more accurate \tctff\ profiles for a large sample of clusters, many of which are actively experiencing thermally unstable cooling.   What differentiates this from preceding studies is careful attention to mass profile measurements within the central galaxy, and careful attention to deprojected temperature and density measurements.  We show that attention to these details are essential in order to test thermal instability and feedback models.  We conclude that the role of local acceleration as captured by the minimum value of \tctff\ is far less clear than has been previously understood. 

The paper is arranged as follows.  We describe our sample in Section \ref{Section:Sample}, and data reduction in Section \ref{Section:DataReduction}.  In Section \ref{Section:Results} we present thermodynamic and mass profiles for our clusters.  Section \ref{Section:Discussion} discusses density and entropy distributions before in Section \ref{Section:CoolingInstabilities} we investigate what causes the onset of thermally unstable gas cooling.  Finally we discuss the possibility of a floor rather than a clear minimum in \tctff\ profiles in Section \ref{Section:tctff_floor} before drawing conclusions in Section \ref{Section:Conclusions}.  Throughout this paper we have assumed a standard $\Lambda$CDM cosmology with: $\Omega_{m}$=0.3, $\Omega_{\Lambda}$=0.7, $H_{0}$=70 km~s$^{-1}$~Mpc$^{-1}$.

\section{Sample Selection} \label{Section:Sample}

\begin{figure}
	\includegraphics[width=0.95\columnwidth]{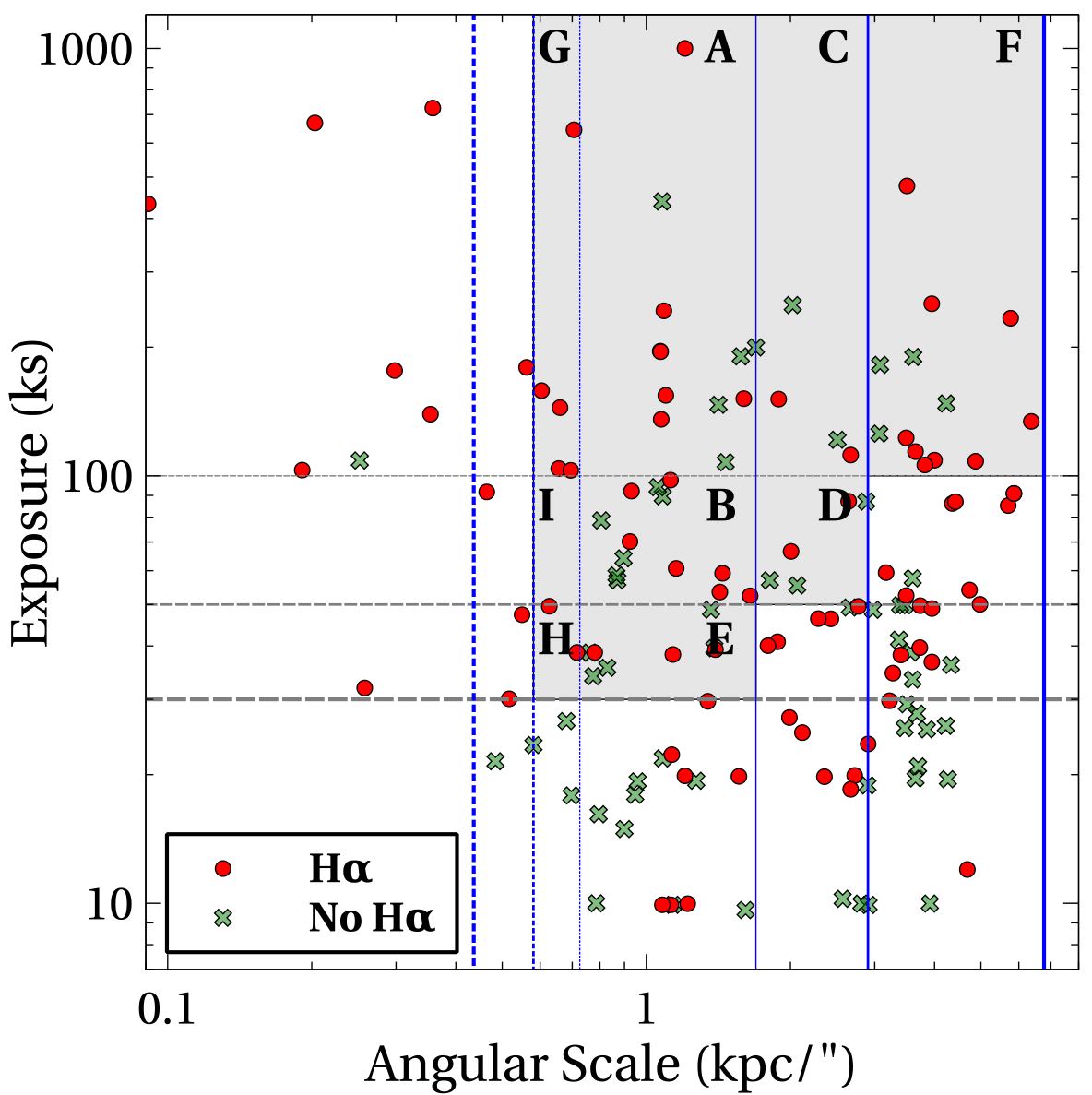}
    \caption{Distribution of the {\it Chandra} exposures available for our parent sample as a function of angular scale on the sky at each cluster's redshift.  Clusters whose BCGs exhibit H$\alpha$ are shown as red circles, those without as green crosses.  The three dashed vertical lines represent limits at which 300, 400, and 500~kpc (left to right) can be recovered for typical cluster placement on the ACIS-I array.  The vertical solid lines show the limits at which 3, 2, and 1 (left to right) annular temperature measurements are reasonable within 10~kpc.  Horizontal lines at 30, 50, and 100~ks total {\it Chandra} exposures are also shown.  This grid creates various regions within which clusters have comparable resolution and depth.  We highlight the regions from which our mass sample have been taken, with labelled boxes corresponding to regions in Table \ref{Table:Observations_Table}.  Further details are given in the text.}
    \label{Figure:Sample_Distribution}
\end{figure}

\subsection{Parent Sample}
We aim to study the role that cluster mass plays in controlling the onset and magnitude of ICM cooling from the hot phase.  To achieve this we require a sample of galaxy clusters that have been observed for tracers of ICM cooling, and which have archival data available from the {\it Chandra} X-ray Observatory (hereforth `{\it Chandra}') online repository.

Initially, we consider the 19 clusters observed for H$\alpha$ in \citet[][]{McDonald10} that also have {\it Chandra} data\footnote{\citet[][]{McDonald10} identify Abell 1837 as having been observed by {\it Chandra}.  This data is not apparent in the archive, although an XMM observation of this target exists.  However, since publication of \citet[][]{McDonald10}, data for Abell 970 has become available and hence the total remains at 19.}.  These targets were selected to cover a large range in cluster richness and cooling rates.  Deep Magellan observations were performed on this sample and hence the presence of multi-phase gas, as traced by H$\alpha$, is known for each of these clusters.  To this sample we add 62 clusters that have been observed for the presence of molecular CO.  These sources comprise the samples of \citet[][]{Edge01} and \citet[][]{Salome03}, in addition to a number of clusters that were observed since the publication of those papers (Edge, {\it private communication}, see also Paper II).  CO is a sensitive tracer of molecular gas and so the presence of significant cooling within these clusters is revealed by the detection, or not, of these lines (Paper II).  The majority of these clusters have also been observed for H$\alpha$ \cite[][]{Crawford99,Cavagnolo09,Rawle12}.

Basing our selection on clusters that have been observed for either CO or H$\alpha$, and which have been observed by {\it Chandra}, naturally biases us towards likely cool-core clusters.  We therefore matched the 81 clusters within this initial sample to \citet[][]{Hogan15a} and the ACCEPT database \citep[][]{Cavagnolo09} to see how many had previously been flagged as likely cool-core clusters due to their BCGs exhibiting optical emission lines \citep[see][]{Crawford99,Cavagnolo08}.  Only 11 of the 81 were expected to be non cool-cores using this proxy.  To ensure a well-sampled range of central cooling time we therefore added the 75 clusters from the ACCEPT database that are tagged as having been observed for H$\alpha$ and which were not already included in our sample.  Of these 75 clusters, 16 had H$\alpha$ detections.

Our parent sample therefore consists of 156 clusters,  of which 86 are expected to contain cool-cores.  The H$\alpha$ coverage of these clusters is heterogeneous, and the presence of these lines is not a perfect indicator for the dynamic state of the cluster \cite[][]{Cavagnolo08}.  However, by selecting our sample to have roughly equal numbers of line-emitting and non line-emitting BCGs we should ensure a wide range of central cooling times are sampled.

During major mergers the dynamically dominant dark matter component of a cluster can become offset from the luminous X-ray ICM (e.g. the `Bullet Cluster') making X-ray derived mass indicators unsuitable \cite[][]{Markevitch02,Clowe06}.  A number of clusters were therefore excluded as a result of clear major merger activity -- A520 \cite[the ``train-wreck'' cluster,][]{Markevitch05}, A115 \cite[][]{Gutierrez05}, A2146 \cite[][]{Russell10,White15}, A754 \cite[][]{Henry95,Macario11}.  Additionally A3158 is a late stage merger \cite[][]{Wang10} and its BCG is positioned on a chip-gap in the ACIS-I array, hence it is also removed from the sample.  A further five sources were removed due to having unsuitable data (chip-placement, etc.).

\subsection{Mass Sample} \label{Section:Mass_Sample}

Thermally unstable cooling in cool-core clusters is typically confined to the central few tens of kpc, as shown by ALMA observations of cold gas \cite[e.g.][]{Russell14,McNamara14,Russell16,Tremblay16,Vantyghem16}.  Furthermore the \tctff\ minimum is usually reported to occur at cluster-centric radii of 5--20~kpc \cite[e.g.][]{Gaspari12,Voit15b,Hogan17}.  High resolution measurements are therefore required to constrain the minimum value of \tctff.  In \citet[][]{Hogan17} we demonstrated the importance of resolving these inner regions and deprojecting both density and temperature.  This is one of the main contributing factors to the difference between our measured thermodynamic properties and those previously reported.  However, the \textsc{clmass} models used to fit cluster mass \cite[][see Section \ref{Section:Mass_Fitting}]{Nulsen10} work best when the full extent of the cluster X-ray atmosphere is sampled.  This places opposing redshift constraints on our sample -- clusters must be close enough to sufficiently resolve the central $\sim$10~kpc (criterion 1) whilst not being so close that their angular extent becomes greater than that observable by {\it Chandra} (criterion 2).

Figure \ref{Figure:Sample_Distribution} shows the distribution of available {\it Chandra} exposure times for our parent sample as a function of the angular scale at each cluster's redshift.  Angular scale is plotted rather than redshift to allow a more direct view of resolvable scales.  It was found in \citet[][]{Hogan17} that taking a minimum circular annulus of radius 3$\times$0.492 arcsec pixels (roughly equivalent to the {\it Chandra} resolution), and then extending the radius of each successive annulus by one pixel (i.e. widths of 3, 4, 5, etc. pixels) provided good spatial sampling whilst ensuring more successful deprojection.  Using these resolution-based annuli as a guide,  we show on Figure \ref{Figure:Sample_Distribution} three vertical solid lines.  The left-most line shows the angular scale at which the three smallest annuli would fall within 10~kpc.  The middle line shows where the two innermost annuli would cover 10~kpc radially, and the right-most line where the innermost annulus alone would cover 10~kpc.  These lines act only as a rough guide since count rate and the presence (or not) of an AGN will place additional constraints on how many radial bins can provide useable spectra in the cluster center.

The maximum cluster field-of-view is difficult to constrain as it depends on the exact observational set-up used as well as the position of the cluster on the ACIS array.  Multiple pointings of a single cluster can also change the available scale.  As an approximate guide we assume that a circular region of $\sim690''$ radius is recoverable -- roughly equivalent to the maximum extent of the ACIS-I array.  Three vertical dotted lines are shown on Figure \ref{Figure:Sample_Distribution}.  These correspond to the angular scale (redshift) at which this maximum angular size corresponds to a recoverable physical scale of radius 300 (left), 400 (middle), and 500~kpc (right).  

As well as spatial constraints, we also desire adequate counts to extract suitable spectra.  We therefore finally show three horizonal dashed lines on Figure \ref{Figure:Sample_Distribution} at raw {\it Chandra} exposure times of 30, 50, and 100~ks.  The count-rate of each observation could equivalently be used here -- though these can be affected by strong point sources and substructures.  Count rate is also likely to disfavor non cool-core clusters, so raw exposure is used as a proxy for depth of observation.

The various constraints plotted on Figure \ref{Figure:Sample_Distribution} create a grid of 9 regions that we label A--I.  Note that region I contains no sources.  The lone source in region H (A2634) was from the sample observed for CO but was found to be an isolated elliptical and thus removed from the sample.  The 56 clusters within the shaded region provide a reasonable compromise between physical resolution and recoverable angular scales whilst having the deepest data.  Of these clusters, 33 are line-emitting.  These clusters constitute our mass sample and are listed in Table \ref{Table:Observations_Table}.

\section{Data Reduction} \label{Section:DataReduction}

Data reduction was performed using \textsc{ciao} version 4.7 with CALDB version 4.6.7 \cite[][]{Fruscione06} following the methods described in \citet[][]{Hogan17}.  A brief outline is given here.

Available imaging data were downloaded from the online {\it Chandra} repository.  Level-1 events were filtered and reprocessed to correct for charge transfer inefficiencies and time dependent gains. \textsc{vfaint} mode was used for more accurate filtering when available.  The \textsc{lc\_clean} script by M. Markevitch was used to remove periods suffering from background flares.  In instances of multiple pointings to a single source, the separate \textsc{obsid}s were reprojected to a common position.  Blank-sky backgrounds were processed in an identical manner for each observation and normalized to the corresponding 9.5--12.0~keV flux.  A background and PSF corrected 0.5--7.0~keV image was created for each cluster.  This was used to identify point-sources and clearly non-equilibrium ICM structures such as cavities and filaments.  These were masked out from subsequent analysis.  The strucure identification was done using the \textsc{wavdetect} \cite[][]{Freeman02} algorithm supplemented by manual inspection in DS9 \cite[][]{Joye03}.

\subsection{Spectral Extraction}

As mentioned in Section \ref{Section:Mass_Sample} \cite[see also][]{Hogan17}, we require deprojected densities and temperatures for robust determination of \tc\ and related quantities.  Retaining sufficient counts after deprojection to measure temperature often requires that spectra be extracted from large regions.  However, we ideally want to sample \tctff\ at altitudes $\lesssim$10~kpc and so small central annular regions are desirable.  These opposing requirements lead us to extract two separate sets of concentric circular annuli for each cluster.  The first is a set of 16 annuli identical to those used and described in \citet[][]{Hogan17}, with radii dictated solely by angular resolution (see also Section \ref{Section:Mass_Sample}).  These annuli provide the highest reasonable radial sampling but for many clusters the small central annular regions may contain insufficient counts for successful spectral fitting.  This is particularly prevalent in non cool-cores with hot diffuse central ICMs.  We therefore extract a second set of annuli for each cluster, where the central region is defined to include a set number of counts.  The exact number of counts per annulus varies by cluster.  This limit again requires a compromise -- too few counts and there is less likelihood of successful fitting, too many counts and the radial binning becomes uselessly large and/or there are too few bins to recover a practical profile.

There is no strict limit on the number of counts required to successfully fit (de)projected temperatures, although hotter clusters typically need more.  We set 3000 counts as the hard minimum required per spectral region.  A single central region of radius $<$10~kpc with more counts is preferred to multiple regions $<$10~kpc each with fewer counts.  Practically, it was found that all expected cool-core clusters in our sample could have at least one annulus within $\sim$10~kpc containing $\gtrsim$4500 counts,  Often a central annulus $<$10~kpc and with $\gtrsim$8000 counts was possible.  Non line-emitting clusters have lower surface brightness peaks thus less counts centrally.  Amongst these, larger central annuli (radii $\sim$20-30kpc) were used in a number of cases to ensure that our minimum count limit of 3000 was not breached.  However these clusters' lack of line emission shows their ICM to not be condensing and they are thus expected to have high central cooling times. The loss of radial resolution of their various parameter profiles (e.g. \tc, entropy) is therefore acceptable and should not impact our results.

Two sets of spectra were therefore extracted for each cluster; one from each of the two sets of annular regions described above.  The \textsc{ciao} tasks \textsc{mkacisrmf} and \textsc{mkwarf} were respectively used to create individual redistribution matrix files (RMFs) and auxillary response files (ARFs) for each spectrum, and exposure maps created to correct each observation for lost area.  Spectra were binned to ensure 30 counts per channel.  In instances of clusters having multiple observations spectra were extracted and treated separately for each \textsc{obsid}.  Since these could be separated greatly in time they were not summed but instead later loaded and fitted simultaneously within the modelling package \textsc{xspec} \cite[][]{Arnaud96}.

\section{Results and Mass Profiles} \label{Section:Results}

\subsection{Thermodynamic Properties of the ICM}  \label{Section:ICMProperties}

\begin{figure*}    
  \centering
    \subfigure{\includegraphics[width=5cm]{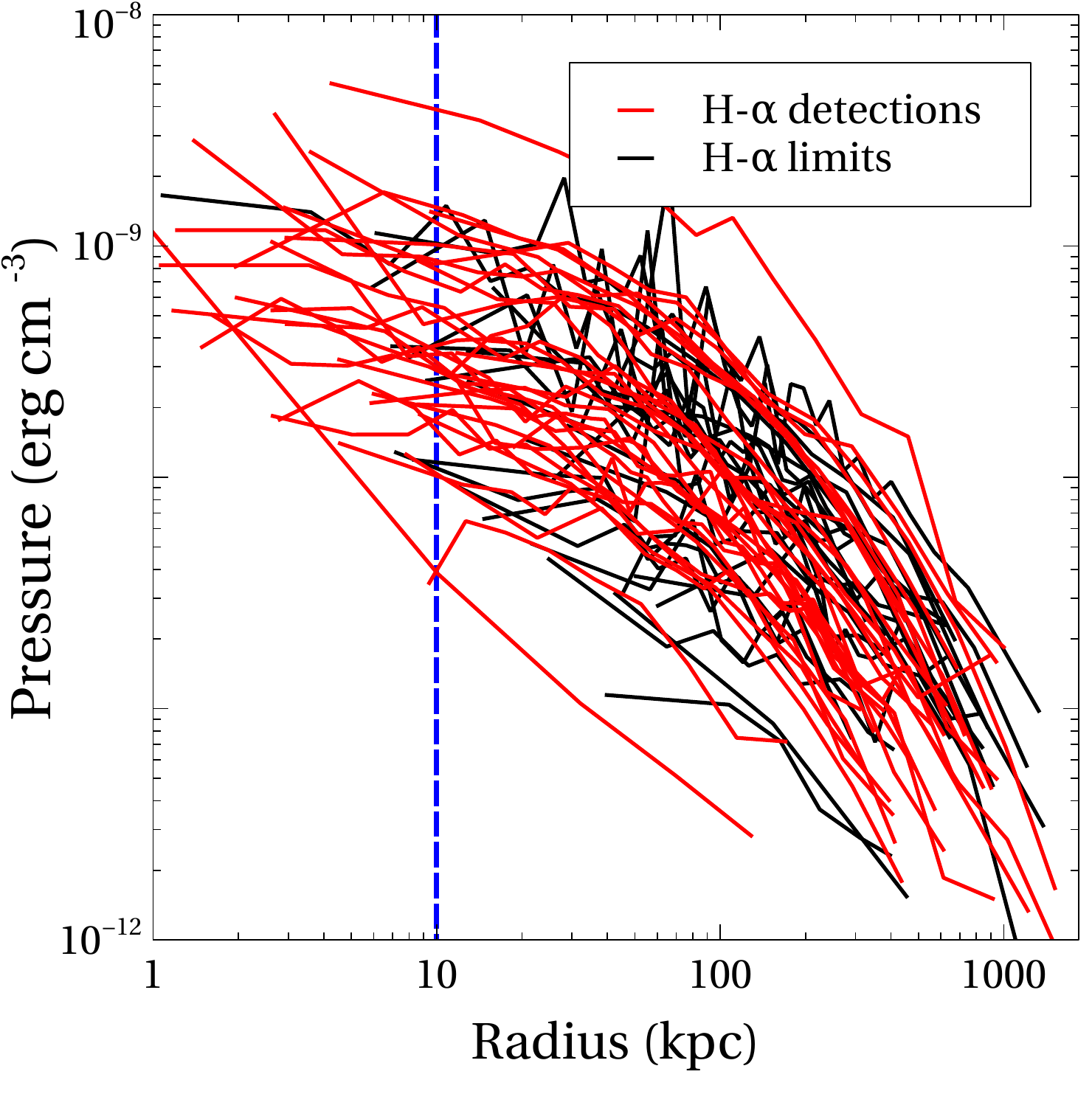}}
    \subfigure{\includegraphics[width=5cm]{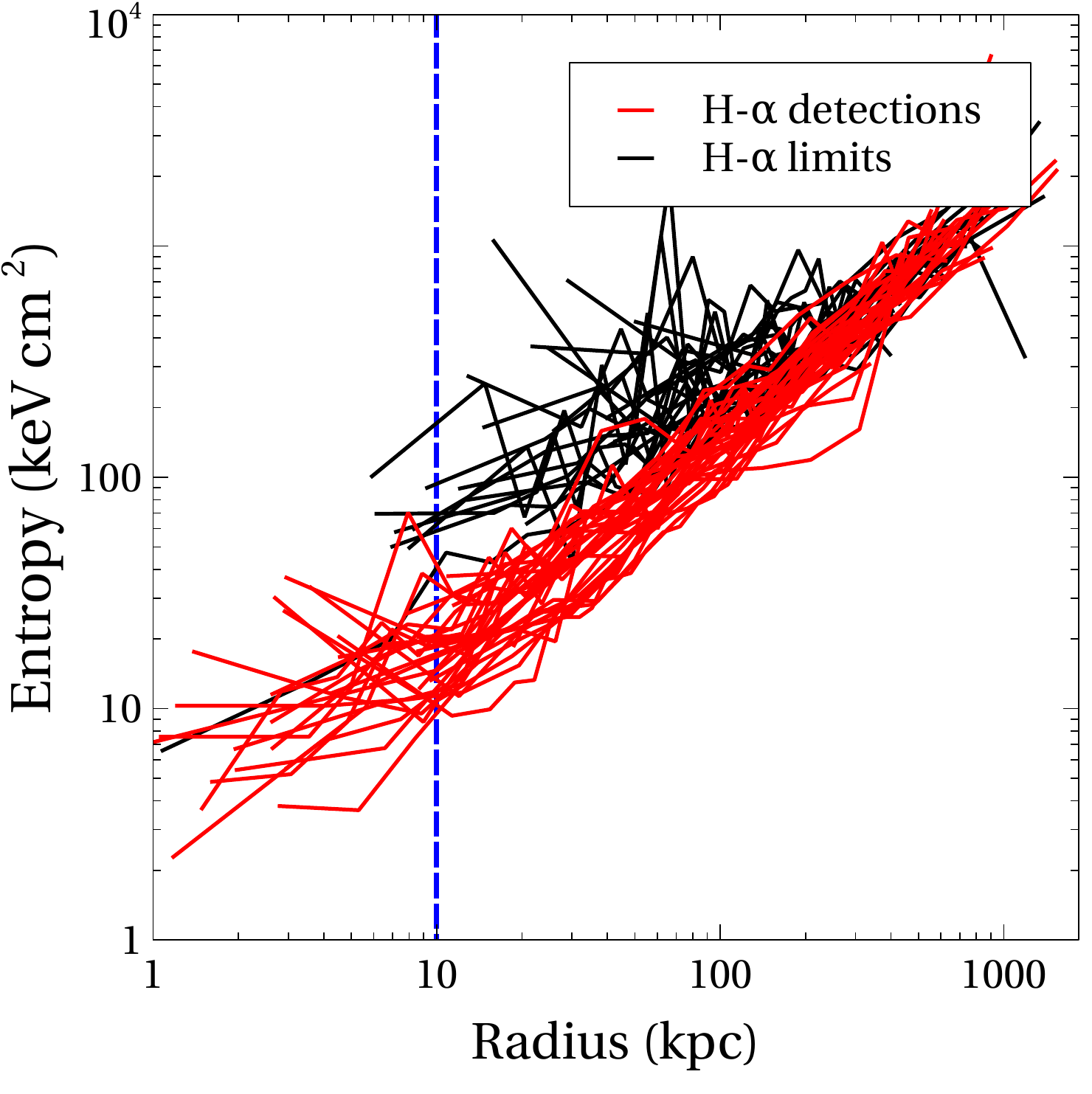}}
    \subfigure{\includegraphics[width=5cm]{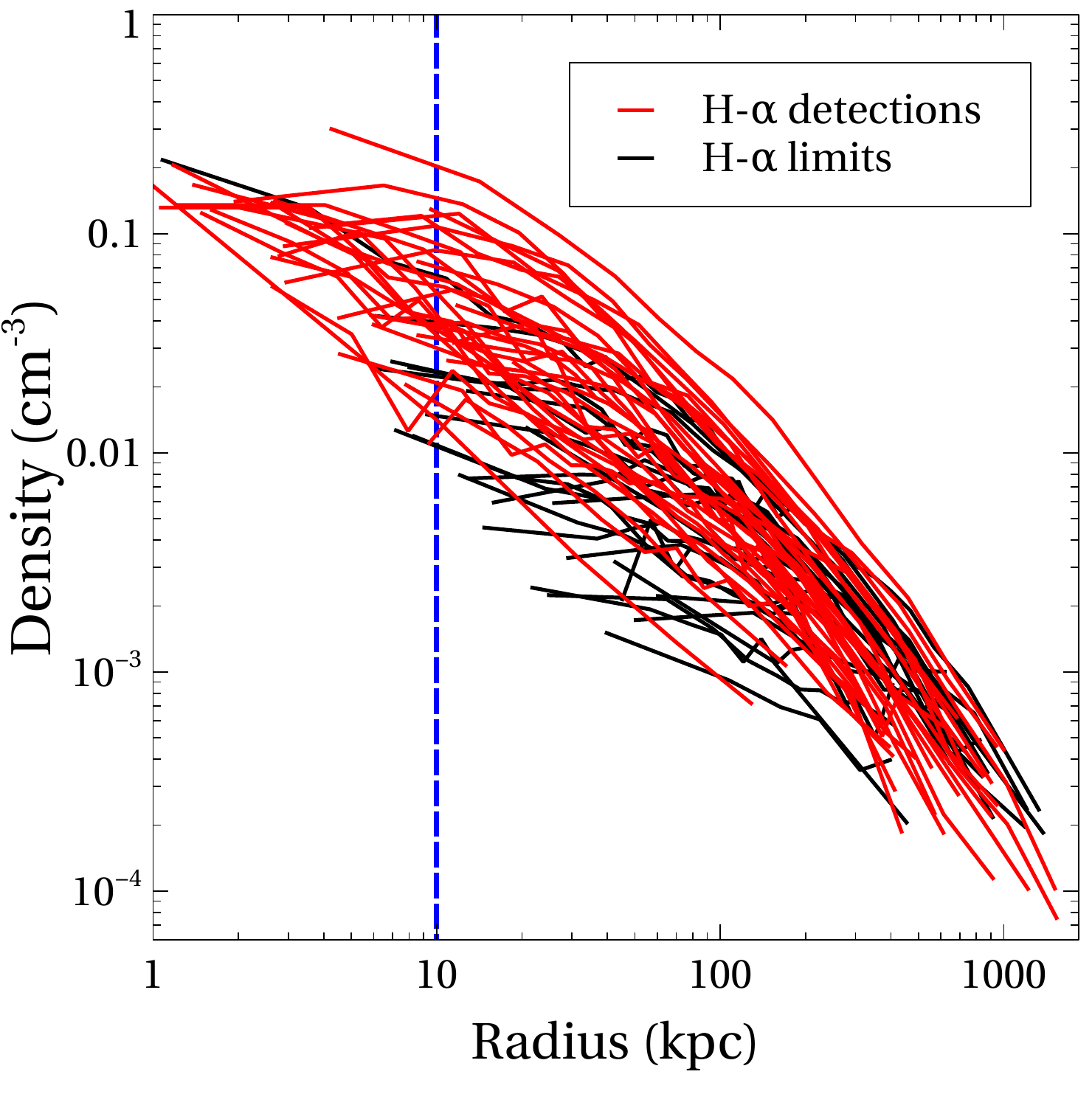}}
  \caption{Deprojected pressure, entropy, and density profiles for our sample, colorised by presence or not of nebular H$\alpha$ emission.  Note that whilst the uncertainty of deprojecting the more diffuse NLE clusters is apparent in these plots, clear trends can still be recovered.  In particular note the lower central entropies of LE clusters (see Section \ref{Section:EntropyProfiles}).  Error bars have been removed for clarity.}
 \label{Figure:DeprojectedProfiles_DensityPressureEntropy}
\end{figure*}

\subsubsection{Projected Profiles}  \label{Section:ProjectedProfiles}

We initially calculate projected thermodynamic profiles, which provide a base from which the effects of deprojection on the final values of \tctff\ can be understood.
 
The extracted spectra and corresponding response files were loaded into \textsc{xspec} version 12.8.2 for spectral fitting \cite[][]{Arnaud96}. We fitted the spectra with an absorbed single temperature (\textsc{phabs}*\textsc{mekal}) model \cite[][]{Mewe85,Balucinska-Church92,Liedahl95}, which was found in \citet[][]{Hogan17} to give a good description of the ICM across our radii of interest.  Solar abundances were set to those of \citet[][]{Anders89}, and line of sight galactic extinctions were frozen to values taken from the LAB survey \cite[][]{Kalberla05}, unless the best fit was found to be significantly different.  For each cluster, preference was given to the set of annuli with finer central radial sampling.  However, sometimes this resolution-based binning left too few counts in the central few spectra to obtain convergent fits.  In these cases we instead used the set of count-based annuli to recover profiles for various ICM properties.  Regardless of which set of annuli were used, convergent fits were sometimes not possible for the smallest radial bins -- most often in non cool-core clusters.  In these instances, multiple central regions could be combined.  However, we opt against this since fits over very large central regions where the temperature may be rapidly changing can bias high subsequent measures of central cooling and entropy \cite[][]{Panagoulia14a,Hogan17}.  Instead we truncate our subsequent profiles at the smallest radial annulus to which a stable spectral fit is recovered.

Temperatures and normalizations from the fitted models were used to derive projected electron number densities n$_{\rm e}$
\begin{equation}
 n_{\rm e}~=~D_{\rm A}(1+z)10^{7}~\sqrt[]{\frac{N~4~\pi~1.2}{V}}
\end{equation}
where N is the model normalization, D$_{\rm A}$ is the angular distance to the source, and V the volume of a spherical shell bounded by the inner and outer projected annulus edges.  The factor of 1.2 arises from the relative abundances of electron n$_{\rm e}$ to ion $n_{\rm H}$ number density \cite[][]{Anders89}.  Cooling times were calculated using 
\begin{equation}
 t_{\rm cool}~=~\frac{3~P}{2~n_{\rm e}~n_{H}~\Lambda(Z,T)} = \frac{3PV}{2L_{X}}
\end{equation}
where P is pressure (${\rm P}=2n_{\rm e}k_{B}T$), and $\Lambda({\rm Z,T})$ the cooling function for gas at a specific abundance Z and temperature T.  The bolometric X-ray luminosity L$_{\rm X}$ is found by integrating the fitted model between 0.1--100~keV.  We finally calculate the specific entropy ($K~=~kT~n_{\rm e}^{-2/3}$) of the ICM, which provides an imprint of the thermal history of a cluster \cite[][]{Panagoulia14a}.

\subsubsection{Deprojected Profiles} \label{Section:Deprojected_and_Density_Stuff}

Spectra extracted from the inner regions of a cluster are contaminated by projected emission from higher altitudes.  An accurate measure of the inner cluster properties therefore requires deprojection of the spectra to remove this superposed emission.  The model independent \textsc{dsdeproj} routine is used to deproject our spectra (\citealt[][]{Russell08}, also see \citealt[][]{Sanders07,Sanders08}). Absorbed single temperature  (\textsc{phabs}*\textsc{mekal}) models are fitted to the deprojected spectra, as for the projected spectra.  

Deprojected density, pressure, and entropy profiles are presented in Figure \ref{Figure:DeprojectedProfiles_DensityPressureEntropy}, colorised by the detection or not of H$\alpha$ emission \cite[][]{Crawford99,Cavagnolo09,Rawle12}.  Deprojected densities are typically 10--50\% lower than the equivalent projected values.

\subsection{Mass Profiles}  \label{Section:Mass_Fitting}
A major source of uncertainty when comparing cooling models to data concerns the difficulty of observationally measuring the dynamical times of the cooling gas.  The simplest dynamical timescale, the free-fall time \tff, relies only on the enclosed mass and is commonly approximated as 

\begin{equation} \label{Equation:tff}
t_{\rm ff}~=~\sqrt\frac{2~r}{g}
\end{equation}

\noindent \cite[e.g.][]{Gaspari12,McDonald15}, where $g$ is the standard gravitational acceleration.  Free-fall time is difficult to measure for any sizeable sample of galaxy clusters, particularly at the low altitudes where it is believed to be most important in the context of cooling instabilities \cite[e.g.][]{McCourt12,Sharma12b}. Hydrostatic mass estimates at $\lesssim$10~kpc are possible for only the most nearby clusters \cite[e.g. M87, see][]{Romanowsky01,Russell15}.  Stellar velocity dispersions can be used to infer the enclosed gravitating mass within the central galaxy, though are only available for a minority of BCGs.  In \citet[][]{Hogan17} we presented a method for calculating cluster mass profiles across a wide radial range. This is done for our current sample, and the mass profiles subsequently used to calculate \tff. A brief outline of the method is given here.

\subsubsection{Cluster Mass Profiles}
Our mass profiles contain two components -- an NFW component to account for the majority of the cluster mass on large scales, and an isothermal sphere to account for the stellar mass of the BCG.  We initially obtain isophotal radii r$_{\rm k20}$ and apparent K-band magnitudes m$_{\rm k20}$ from 2MASS \cite[][]{Skrutskie06} for the BCG in each cluster.  These are extinction, evolution, and K-corrected \cite[][]{Poggianti97,Schlegel98}, then converted to enclosed stellar masses within r$_{\rm k20}$ \cite[][]{Bell03,Baldry08}.  An {\em equivalent stellar velocity dispersion} $\sigma_{*}$ is calculated for each of these masses using \citet[][]{Pizzella05}, which describes the isothermal potential

\begin{equation} \label{IsothermalEquation}
\Phi_{\rm iso,c}(r) = \sigma_{*}^{2}\ln(1+(r/r_{\rm I})^{2}) ,
\end{equation}

\noindent where r$_{\rm I}$ is an isothermal scale radius.  This form of isothermal potential is used for numerical reasons.  In practice r$_{\rm I}$ is set to an arbitrarily small but non-zero value ($\ll$1~kpc) so the isothermal potential is equivalent to that of a basic singular isothermal sphere, $ \Phi_{\rm iso}({\rm r}) = 2  \sigma^{2} \ln({\rm r})$, at all radii of interest.  

Based on the methods of \citet[][]{Main17}, we use the \textsc{clmass} \cite[][]{Nulsen10} package of cluster mass mixing models to fit an \textsc{isonfwmass} model to the {\it Chandra} data.  This model combines an isothermal potential (Equation \ref{IsothermalEquation}) with an NFW potential, 

\begin{equation}  \label{Equation:NFWPotential}
\Phi_{\rm NFW}(r) = -4{\pi}G\rho_{0}~r_{\rm s}^{2}~\frac{\ln(1~+~r/r_{\rm s})}{r/r_{\rm s}}
\end{equation}

\noindent where r$_{\rm s}$ is scale radius.  We fix the isothermal potential to that calculated for the stellar component of the BCG, meaning that the remaining cluster mass is fitted with the NFW.  In \citet[][]{Hogan17} we found this method to provide reliable cluster mass estimates from small ($\gtrsim$1~kpc) radii up to R$_{\rm 2500}$.  Model parameters and cluster masses for our sample are presented in Table \ref{Table:ISONFW_fits}.  As a sanity check, total cluster masses (M$_{\rm 2500}$) were compared to other published values where available \cite[e.g.][]{Vikhlinin06,Allen08,Vikhlinin09,Main17}.  Good overall agreement was found.  Notes on some individual clusters can be found in Appendix \ref{Appendix:Notes_on_Individual_Clusters}.

For our sample, the mean of the equivalent stellar velocity dispersions is 268.9$\pm$7.7~km~s$^{-1}$, with a standard deviation of 58.3~km~s$^{-1}$.  As this stellar component is a major contributor to the acceleration at the altitudes where \tctffmin\ is typically found, the large range shows that individually tailored inner mass profiles are required for accurate estimates of the \tctffmin.

\subsubsection{Differences from \citet[][]{Hogan17}}

Our approach here is slightly different from \citet[][]{Hogan17}.  The \textsc{clmass} models contain a switch to allow the inclusion of a $\beta$-model component in the cluster mass profile to account for emission outside of the field of view (also see Section \ref{Section:Mass_Sample}).  Typically this switch is turned off, which gives more stable fits.  In order to avoid underestimating the cluster mass by missing emission beyond the field of view, a mass model with the $\beta$ parameter set free was fitted and the result compared to the original model. An F-test was used to determine whether the $\beta$-model provided a better fit.  We found that a $\beta$-model was justified for only 5 clusters (A133, A401, A1991, A1758, A2052: see Table A2).

A further issue not encountered in \citet[][]{Hogan17} was that four clusters (A665, MACS1347-11, MACS1423+24, and MACS1532+30) are at high enough redshift that their BCGs are undetected in 2MASS.  These are all found in Region F of Figure \ref{Figure:Sample_Distribution} and so typically have only a single region within the innermost region where the isothermal component is expected to dominate ($\lesssim$10~kpc).  We take the mean $\sigma_{*}$ of all clusters in Region F with a 2MASS detected BCG, and adopt this value as an estimate of the isothermal component for these four clusters and then fitted them similarly to the others.

\section{Discussion} \label{Section:Discussion}

\begin{figure}
	\includegraphics[width=\columnwidth]{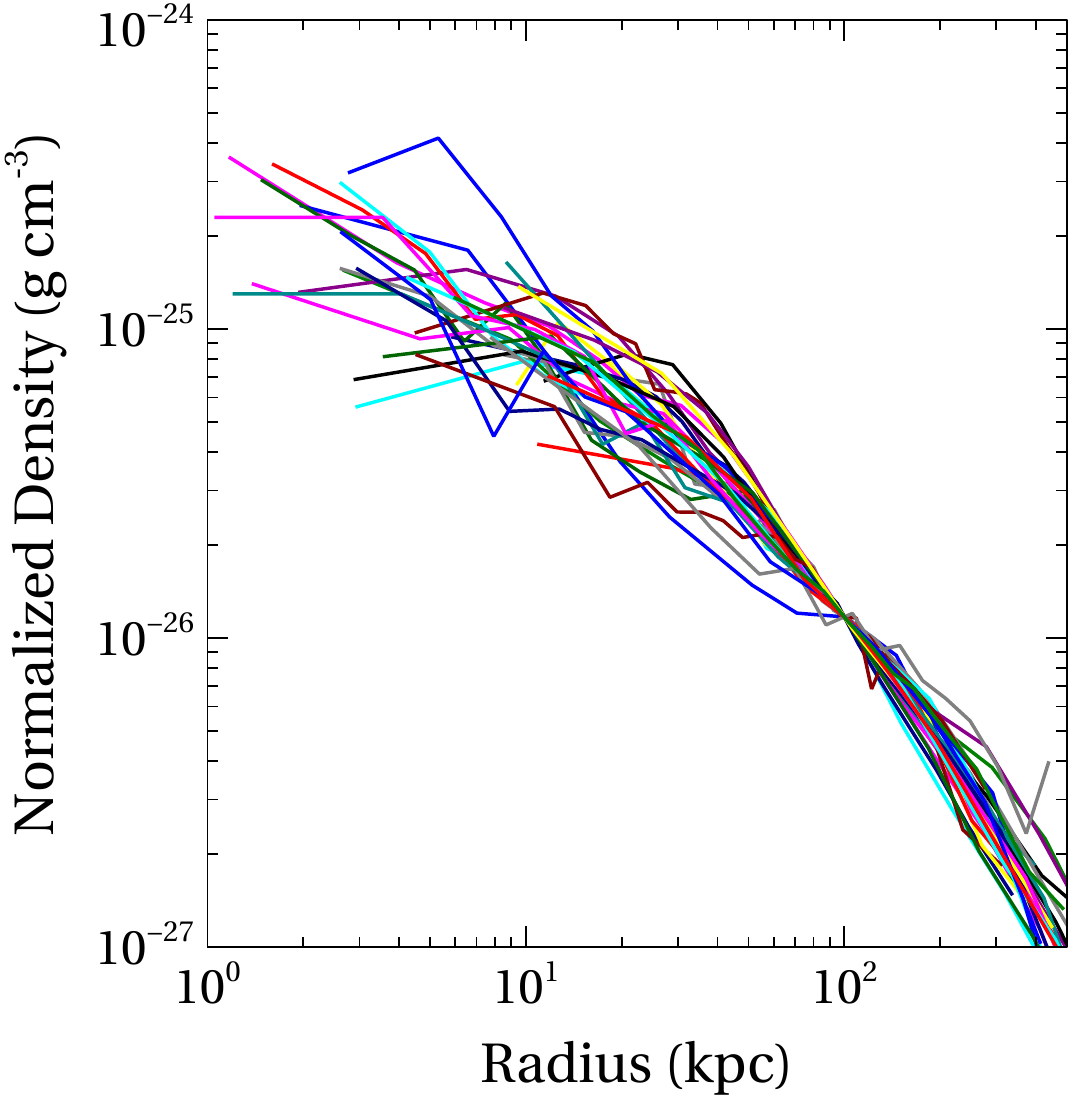}
    \caption{Deprojected density $\rho$ profiles (see also Figure \ref{Figure:DeprojectedProfiles_DensityPressureEntropy}) normalized to median density at 100~kpc for the H$\alpha$ emitting clusters in our sample.  Density is plotted in mass units for ease of comparison with other studies.  Profiles have been randomly assigned colors for presentation purposes.}
    \label{Figure:Normalised_Densities}
\end{figure}

\subsection{Density distribution requires gentle AGN feedback} \label{Section:DensitySwings}


Several features are noteworthy in Figure \ref{Figure:DeprojectedProfiles_DensityPressureEntropy}.  First, deprojection was less successful for clusters without nebular emission (i.e. non cool-cores).  These systems can typically only have their properties traced to higher altitudes because their atmospheres are more diffuse, disturbed, and fainter.  Nevertheless, interesting trends are seen.

The left-hand panel of Figure \ref{Figure:DeprojectedProfiles_DensityPressureEntropy} shows no clear difference between the atmospheric pressures of cooling and non cool-core clusters. However, the entropy profiles of cool-core and non cool-core clusters (middle panel) segregate, confirming the threshold discovered by \citet[][]{Rafferty08} and \citet[][]{Cavagnolo08}. The single H$\alpha$ non-emitting low entropy cluster is A2029, a well known anomaly \cite[see e.g.][]{McNamara16}.  The entropy dichotomy is a consequence of higher central temperatures and lower central densities of non cool-core clusters, and the converse (right-hand panel, Figure \ref{Figure:DeprojectedProfiles_DensityPressureEntropy}).

Significantly, the spread of atmospheric gas density is relatively small at all radii in cool-core clusters (see Figure \ref{Figure:DeprojectedProfiles_DensityPressureEntropy}).  Most of this spread correlates with halo mass.  However, most feedback models indicate that density profiles vary throughout the AGN activity cycle.  To account for the dependence on cluster mass, the density profiles of the H$\alpha$ emitting clusters were renormalized to the median density at 100~kpc, i.e., by a factor of $\rho_{\rm median}(100\rm\ kpc) / \rho(100\rm\ kpc)$.  This normalization further reduces the spread in observed inner densities.  The remaining spread can now be understood as the scatter caused by AGN feedback and other local atmospheric inhomogeneities.

The observed range in central density is strikingly smaller than simulations suggest.  We find no evidence for the large variations in gas density expected if \tctff\ were varying in response to AGN heating and radiative cooling cycles.  In particular, we see no evidence that \tctff\ is rising and falling above and below 10.  Outbursts violent enough to quickly raise \tctff\ back above 10 would cause greater than order-of-magnitude swings in density at the cluster center \cite[$\lesssim$20~kpc, e.g.][]{Sijacki06,Gaspari12,Li15,Prasad15}, which are not observed.  Restricting our analysis to those clusters with H$\alpha$ emission (i.e. cool-core clusters) we find a 10--90th percentile spread in central gas density only factors of 1.2--1.5 wider at 10~kpc than at higher altitudes ($\gtrsim$100~kpc, see Figures \ref{Figure:DeprojectedProfiles_DensityPressureEntropy}) where AGN feedback should be less efficient.  The expected spread in densities at 10~kpc due to heating and cooling cycles is expected to be one to two orders of magnitude greater \cite[e.g.][]{Li15}.  Our sample spans four decades in AGN power, from relatively weak ($\sim$few~10$^{42}$erg~s$^{-1}$) sources such as NGC~5098 and the BCG in A1644, to the most powerful cavity system known, MS0735+7421 ($\sim$10$^{46}$erg~s$^{-1}$).  The small range of central densities shows that central atmospheres do not experience large density swings in response to radio-AGN feedback.  As found by \citet[][]{McNamara16}, radio AGN feedback is a gentle process which levies important and restrictive constraints on jet and feedback models.

\subsection{Cluster Entropy Profiles} \label{Section:EntropyProfiles}

\begin{figure}
	\includegraphics[width=\columnwidth]{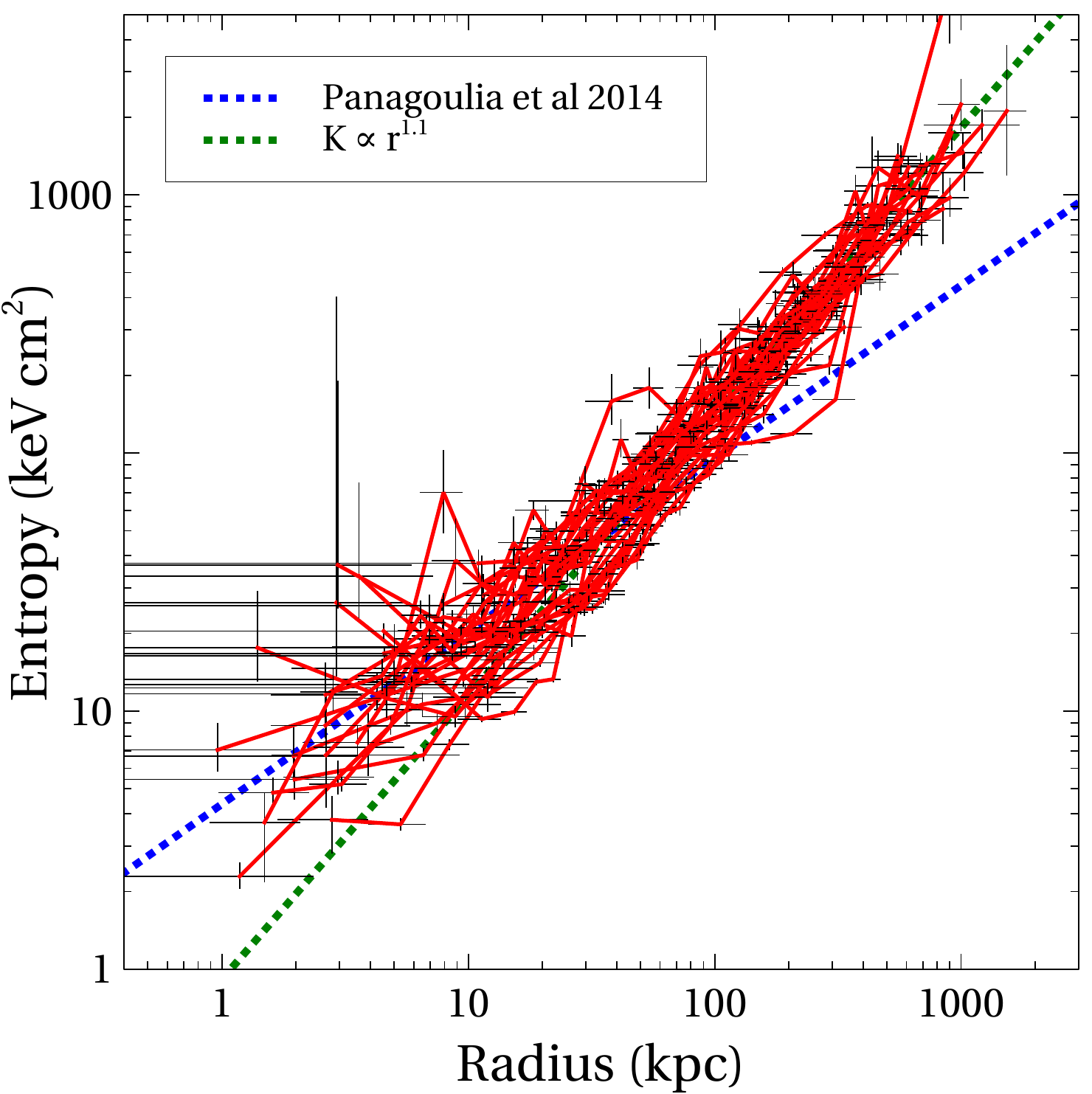}
    \caption{Entropy profiles of all clusters in our sample that display H$\alpha$ emission.  Our inner profiles are in agreement with the average fitted entropy profile of \citet[][]{Panagoulia14a}.  At larger radii we find good agreement with the standard entropy profile power-law shape expected from gravity alone \cite[][]{Tozzi01,Voit05a}.  The dashed green line shown here is taken as a representative cluster, see text.}
    \label{EntropyProfiles}
\end{figure}

\begin{figure}
	\includegraphics[width=\columnwidth]{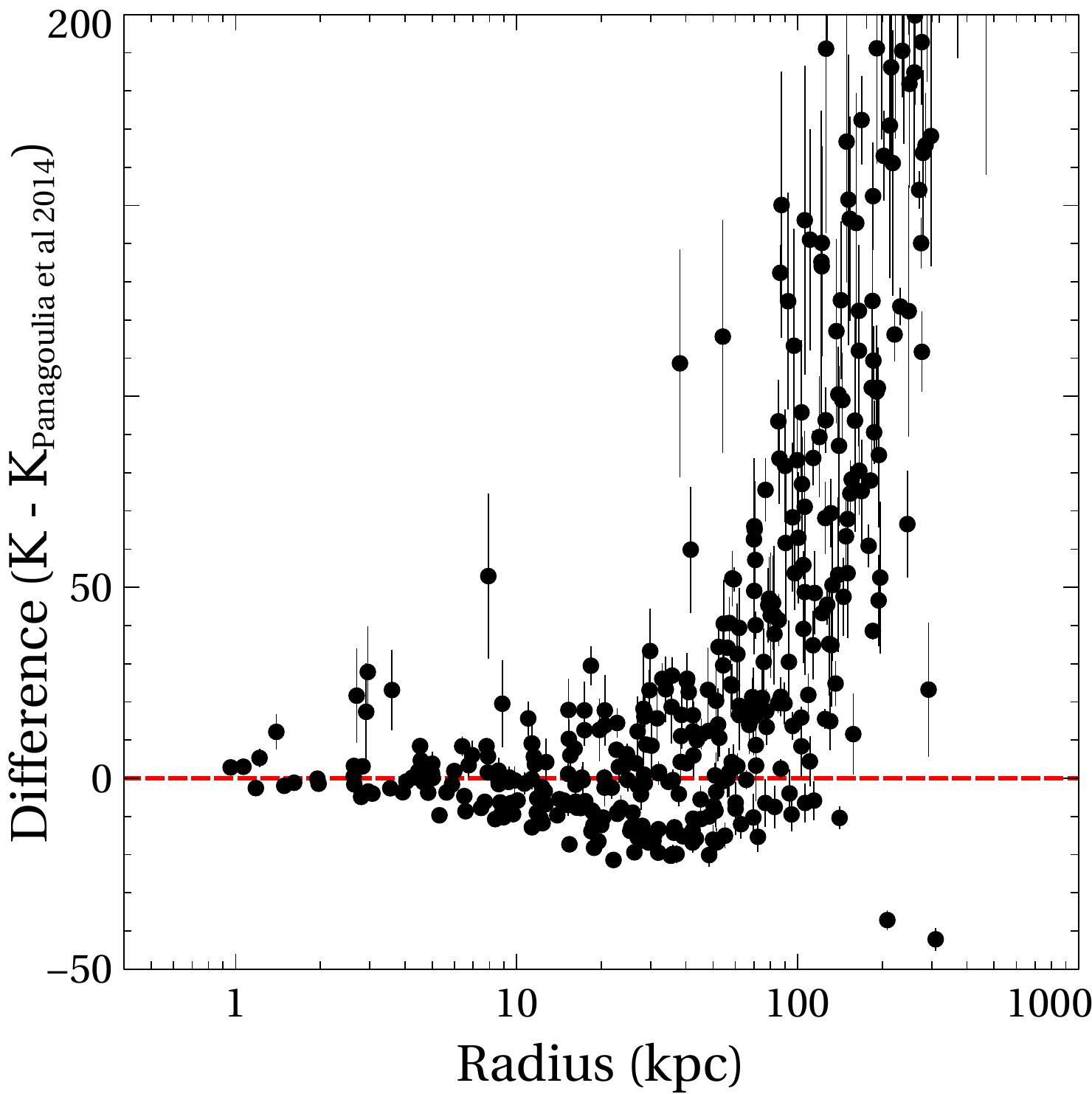}
    \caption{Difference between our calculated entropy profiles and the best fit profile from \citet[][]{Panagoulia14a}, as a function of radius, for clusters in Figure \ref{EntropyProfiles}.  An isentropic core would cause points to lie systematically above the zero-point.  At large radii ($\sim$50~kpc) there is a systematic steepening of the entropy profiles.  The connecting lines for individual clusters have been removed since we are interested in the overall trend here.}
    \label{EntropyDifference}
\end{figure}

The shape of the central entropy profiles in cool-cores is key to understanding thermally unstable cooling \cite[][]{Sharma12b,Voit16}.  For example, in systems where heat is injected centrally, the atmosphere of the core may be almost isentropic \cite[][]{Voit16}.  Beyond the core, convection may stabilise gas against thermal instability, unless low-entropy gas is uplifted allowing it to cool \cite[][]{McNamara16}.

Apparently, gravity alone imprints an entropy power law of the form K$\propto$r$^{1.1}$ \cite[e.g.][]{Tozzi01,Voit05a}.  Other, non-gravitational processes, such as AGN outbursts, may enhance the inner entropy, flattening the profile. Early {\it Chandra} observations indeed found flat or flattening inner entropy profiles in clusters \cite[e.g.][]{David96,Ponman99,David01,Ponman03} which may be fit functionally as $K(r) = K_{0} + K_{100} (r / 100 {\rm kpc})^{\alpha}$ \cite[][]{Donahue05,Donahue06,Cavagnolo08,Cavagnolo09,Voit16}.  This form provides a good approximation to clusters with high central entropy \cite[][]{Cavagnolo09}.  However, it poorly represents cool-cores.  \citet[][]{Panagoulia14a} found that cool-cores are instead characterized by broken power-laws.  The inner 50~kpc are well described by a $K \propto r^{0.67}$ scaling that persists down to at least a few kpc.  Similarly both \citet[][]{Lakhchaura16} and \citet[][]{Hogan17} found cool-core entropy profiles continuing to fall down to small radii.

In Figure \ref{EntropyProfiles} we present fully deprojected entropy profiles for the 33 clusters with central H$\alpha$ emission, 14 of which overlap with \citet[][]{Panagoulia14a}'s sample.  We have overlain \citet[][]{Panagoulia14a}'s mean profile of $K~=~95.4~\times~(r/100~{\rm kpc})^{0.67}$ and it agrees with ours. At large radii the profiles steepen to match the baseline $K \propto r^{1.1}$ power-law scaling expected from gravity alone \cite[e.g.][]{Tozzi01,Voit05a}. The self-similar entropy profile derived by \citet[][]{Voit05a} takes the form  $K = 1.32~\times~K_{200}(r/R_{200})^{1.1}$, where K$_{200}$ is the entropy at R$_{200}$.  Our individual mass profiles become increasingly uncertain beyond R$_{2500}$ \cite[][]{Hogan17}.  To obtain an approximate $K_{\rm 200}$ for our sample we take the calculated R$_{\rm 200}\approx$1240~kpc for a fiducial rich cluster from \citet[][]{Voit05a}, and extrapolate our entropy profile distributions to this radius.  This gives $K_{\rm 200}~\approx~1750~{\rm keV}~{\rm cm}^{2}$, which is the normalization plotted in Figure \ref{EntropyProfiles}.

In Figure \ref{EntropyDifference} we plot the difference between our calculated entropy profiles and the best fit profile of \citet[][]{Panagoulia14a} as a function of radius.  The profiles agree below $\sim$50~kpc, down to $\sim$1~kpc, whereas large isentropic cores, if they existed, would raise the points systematically above the zero-point at low radii.  We do find such a rise above 50~kpc, consistent with the index steepening to $K \propto r^{1.1}$ as seen in Figure \ref{EntropyProfiles}.

Further flattening of the entropy profiles is likely on smaller scales than probed here \cite[see e.g.][]{Donahue06}, especially once the acceleration associated with the central SMBH becomes more relevant.  However, we find that down to a resolution limit of $\sim$1~kpc the entropy profiles of our cool-core clusters are consistent with a broken power-law. 

\begin{figure*}    
  \centering
    \subfigure{\includegraphics[width=8cm]{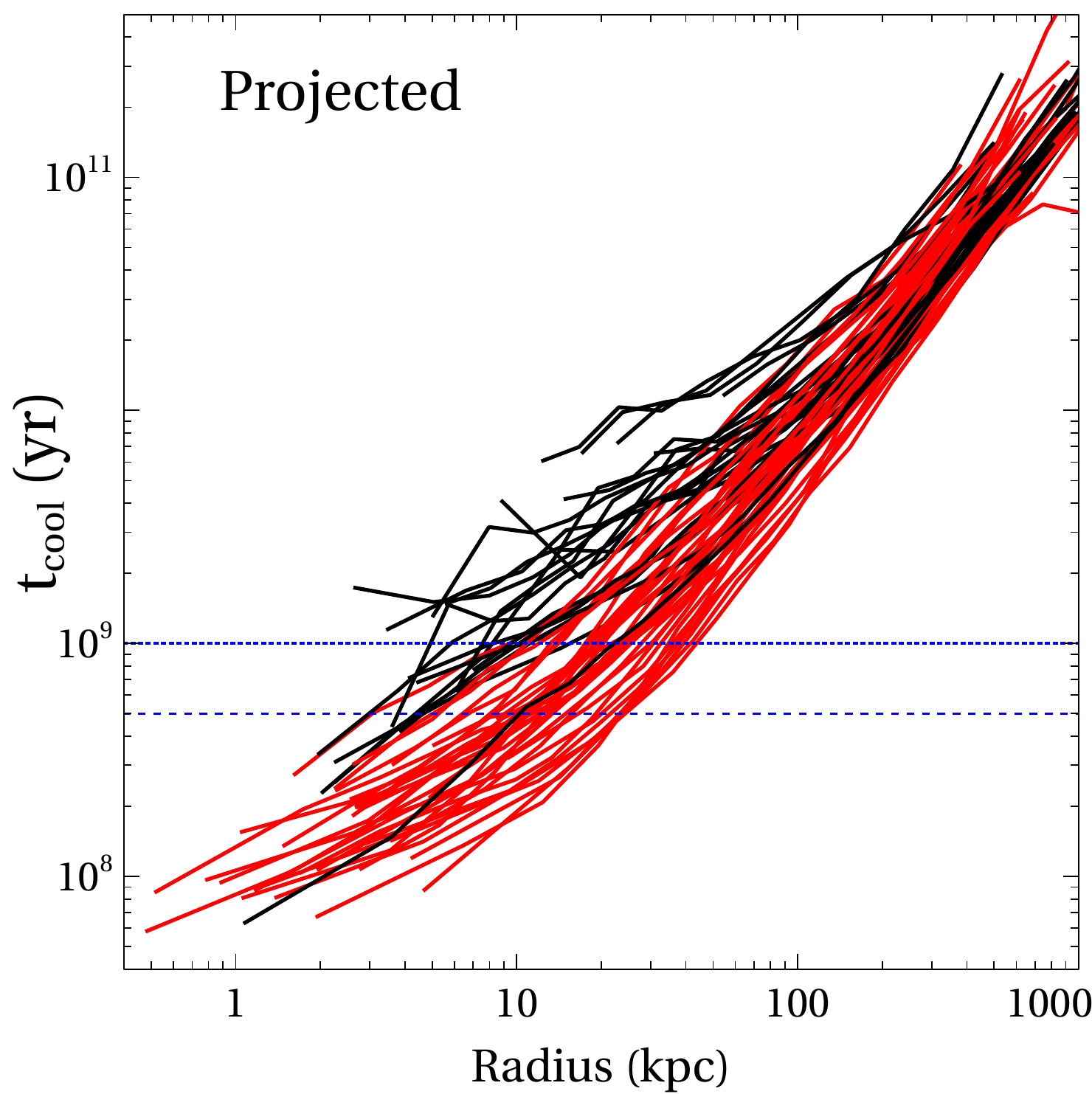}}
    \subfigure{\includegraphics[width=8cm]{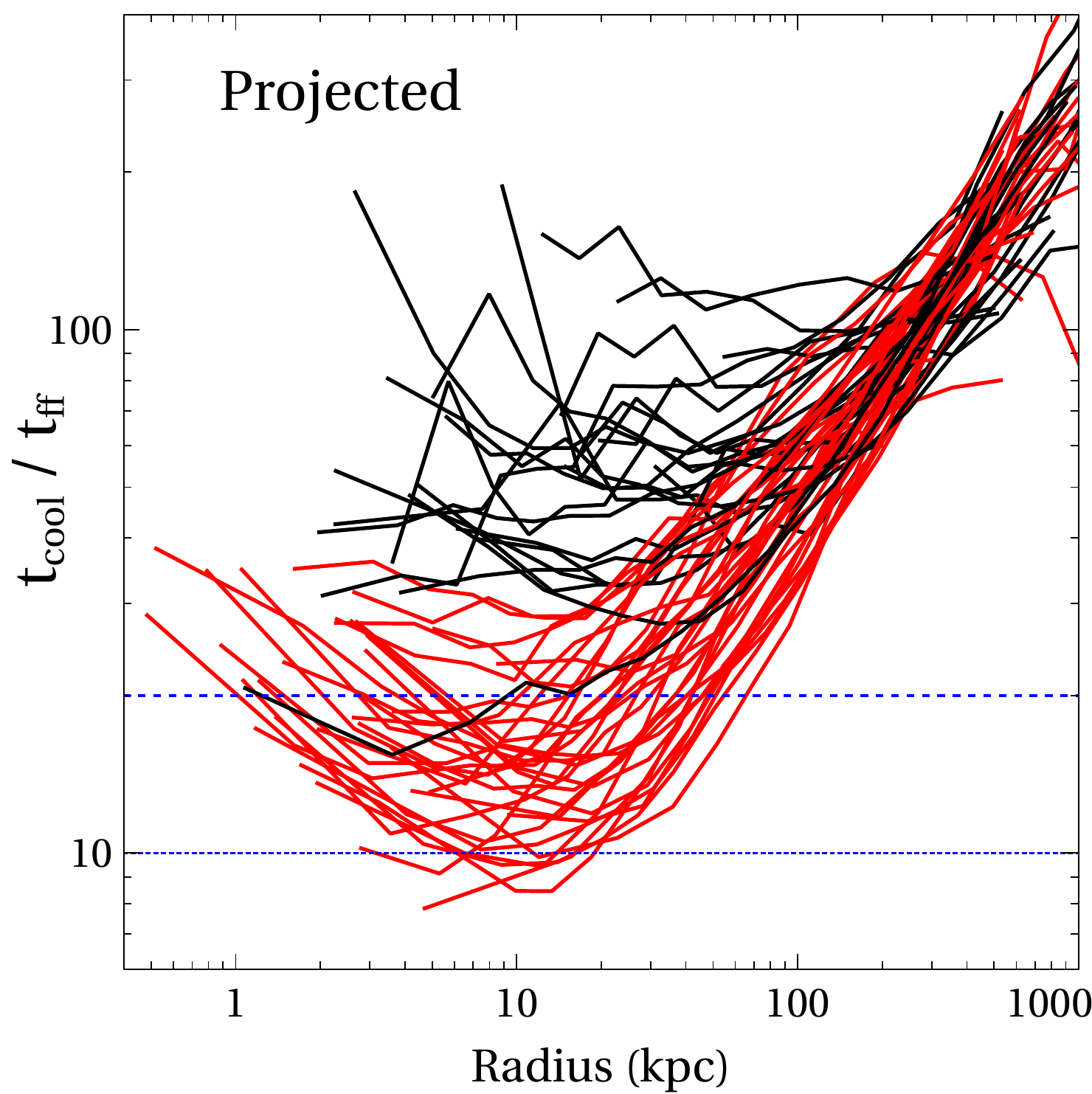}}
  \caption{Projected cooling time (left) and \tctff\ (right), colorised by the presence (red) or absence (black) of nebular emission.  Clusters lacking nebular emission indicative of ongoing gas condensation typically have both higher central \tc\ and \tctff.  Most of the black profiles apparently violating the \tc\ $<$1~Gyr threshold do so only due to projection effects (see Figure \ref{Figure:DeprojectedProfiles_tctctff}).  The single cluster without nebular emission yet with a central cooling time below 1$\times10^{8}~{\rm yrs}$ and a \tctffmin$<$20 is A2029.  Error bars are omitted for clarity.}
 \label{Figure:ProjectedProfiles_tctctff}
\end{figure*}

\section{The Onset of Gas Condensation} \label{Section:CoolingInstabilities}

\begin{figure*}    
  \centering
    \subfigure{\includegraphics[width=8cm]{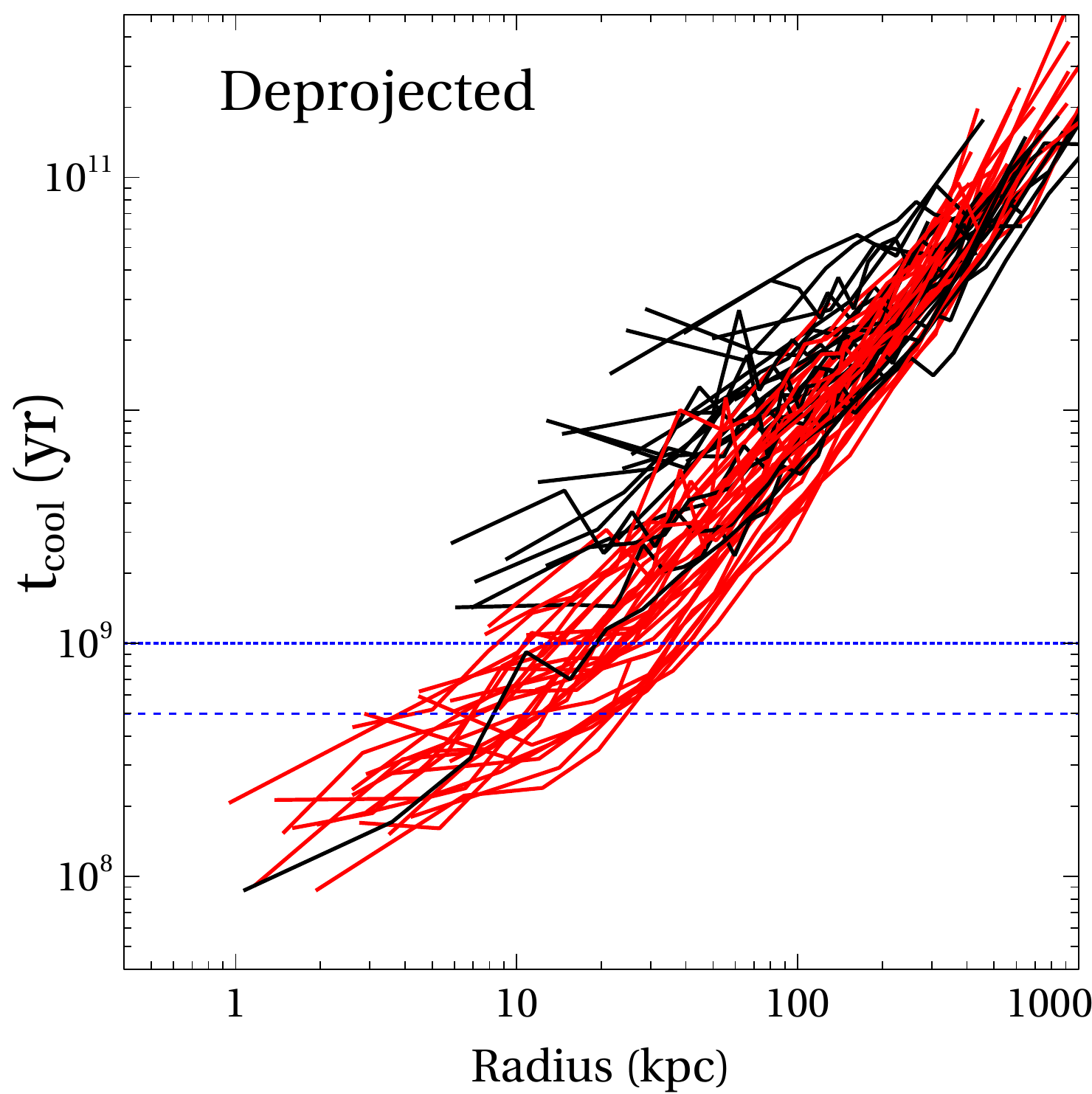}}
    \subfigure{\includegraphics[width=8cm]{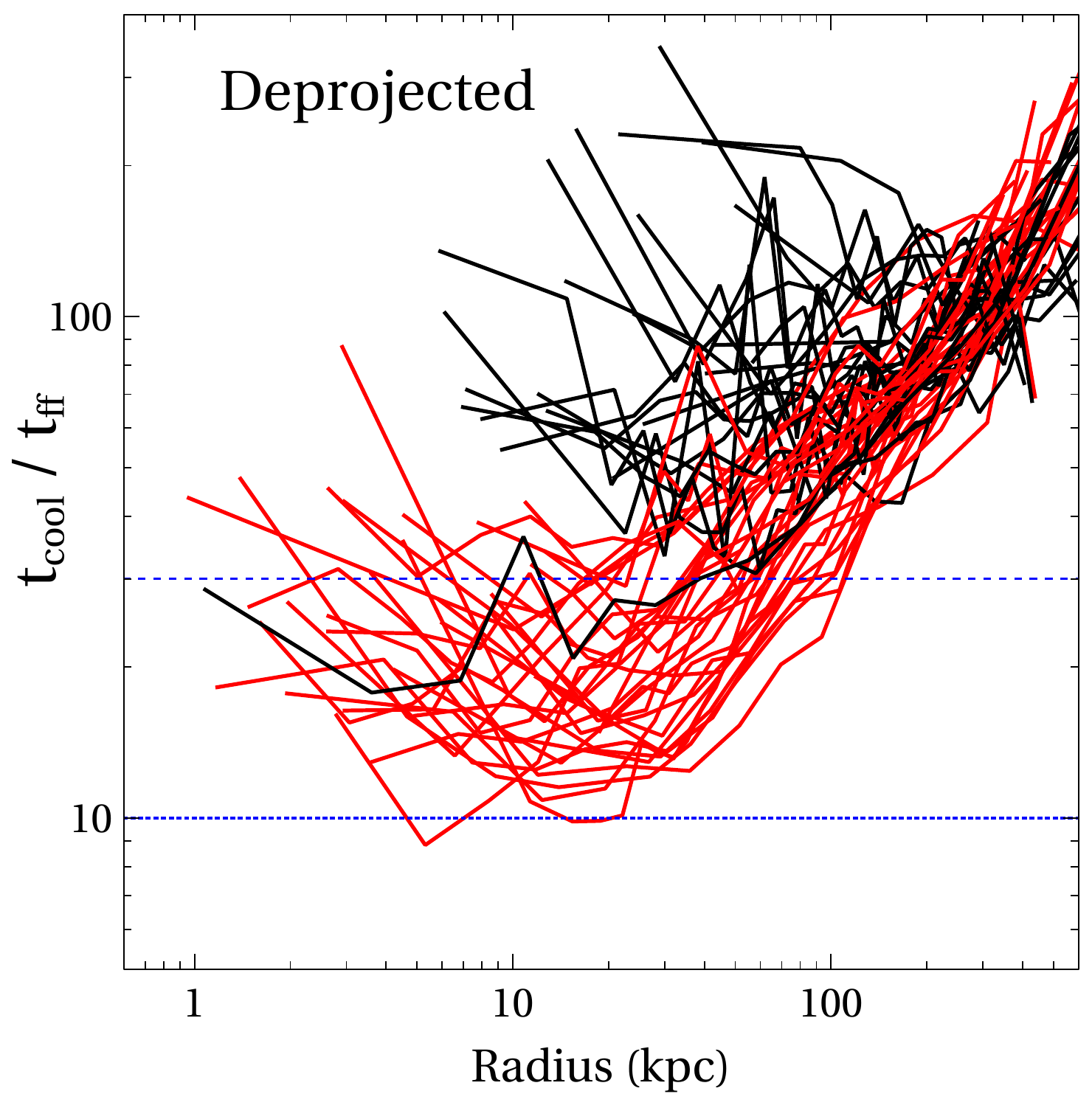}}
  \caption{Deprojected version of Figure \ref{Figure:DeprojectedProfiles_tctctff}. Note that several clusters, mainly the more diffuse systems, had too few counts in their central annuli for successful fitting after deprojection and so these profiles sometimes truncate at larger radii than their projected analogues.  The cooling threshold in the left panel is more sharply defined than in Figure \ref{Figure:ProjectedProfiles_tctctff}, with only the well-known outlier A2029 having \tc\ $<$ 1~Gyr at 10~kpc amongst the non-nebular clusters.  Most cooling clusters have \tctffmin\ in the range 10--30.  Error bars have been omitted to aid clarity.}
 \label{Figure:DeprojectedProfiles_tctctff}
\end{figure*}

The mass profiles derived in Section \ref{Section:Mass_Fitting} were used to calculate \tff\ profiles for each cluster.  A 10\% systematic error on mass was assumed and propagated into \tff\ \cite[see][]{Hogan17}.  Combining these with the atmospheric modelling from Section \ref{Section:ICMProperties} we present projected and deprojected profiles for both \tc\ and \tctff\  (Figures \ref{Figure:ProjectedProfiles_tctctff} and \ref{Figure:DeprojectedProfiles_tctctff} respectively).  We indicate in the left-hand panels of Figures \ref{Figure:ProjectedProfiles_tctctff} and \ref{Figure:DeprojectedProfiles_tctctff} the approximate thresholds of 5$\times10^{8}~{\rm yr}$ and 1.0$\times10^{9}~{\rm yr}$, below which nebular emission and star formation are observed \cite[][]{Cavagnolo08,Rafferty08}.  In the right-hand panels of these figures we indicate \tctff\ threshold values of 10 and 20.

Comparison of the left-hand panels of Figures \ref{Figure:ProjectedProfiles_tctctff} and \ref{Figure:DeprojectedProfiles_tctctff} shows that once projection effects are accounted for, we recover the bimodality between clusters with short and long cooling times.  Abell 2029, with its short \tccentral\  yet no H$\alpha$ emission \cite[][]{McDonald10} is clearly an outlier \cite[discussed extensively in][]{McNamara16}.  The right-hand panels similarly split between line-emitting and non line-emitting clusters, A2029 again being an outlier. The values of \tctffmin\ for objects with central H$\alpha$ emission (red lines) range between 8.8--30.3, with a mean of 16.5 and standard deviation of 5.7.  Only a single cluster lies below \tctff$=10$, but not significantly so (within 1-$\sigma$).  Therefore, deviations below 10 do not occur or are extremely rare (see also Paper II).  With A2029 included (excluded), the non line-emitters (black lines) have \tctffmin\ spanning 17.8 (30.8)--101.4, with a mean of 62.8 (60.8) and standard deviation of 22.8 (21.4).

\begin{figure}
	\includegraphics[width=\columnwidth]{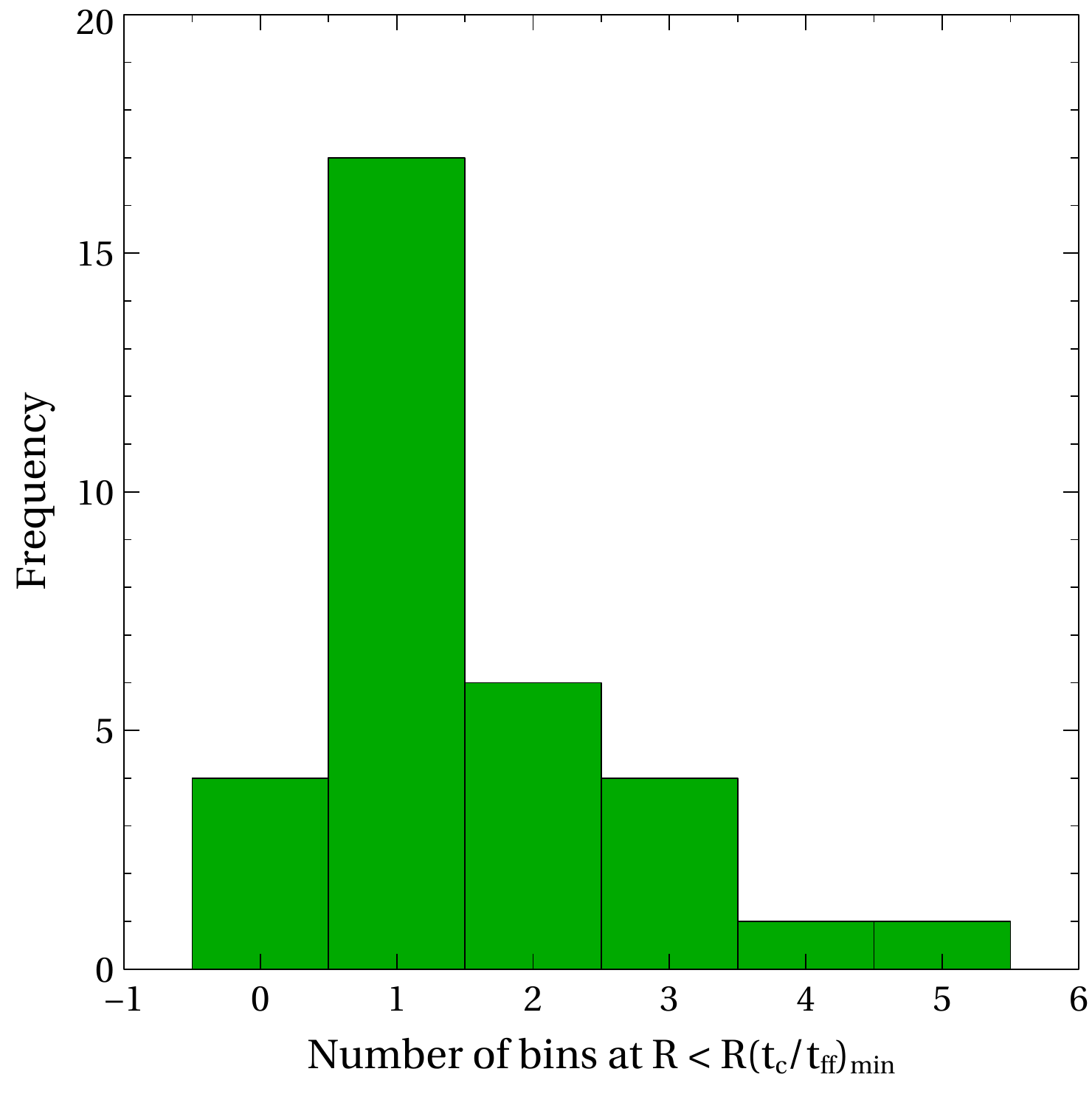}
    \caption{Histogram showing the number of annuli within the annulus having the minimum deprojected \tctff\ (see Figure \ref{Figure:DeprojectedProfiles_tctctff}) for the 33 H$\alpha$ emitting clusters in the sample.  In most cases, there is only a single (noisy) bin at smaller radii, showing that the minima are not well resolved.}
    \label{Figure:minbin}
\end{figure}

Error bars have been excluded on Figures \ref{Figure:ProjectedProfiles_tctctff} and \ref{Figure:DeprojectedProfiles_tctctff} for clarity. However, the right hand panel of Figure \ref{Figure:DeprojectedProfiles_tctctff} suggests that the location of the \tctff\ minimum is noisy. Figure \ref{Figure:minbin} shows a histogram of the number of annuli at radii below that in which the minimum \tctff\ is measured, for all 33 H$\alpha$ emitting clusters.  This implies that \Rtctffmin\ values are usually poorly resolved (in all studies thus far) thus are unreliable (see also Section \ref{Section:tctff_floor}).

\subsection{Thresholds in \tc\ and \tctff} \label{subsection:Thresholds}

\begin{figure*}    
  \centering   
    \subfigure{\includegraphics[width=18.5cm]{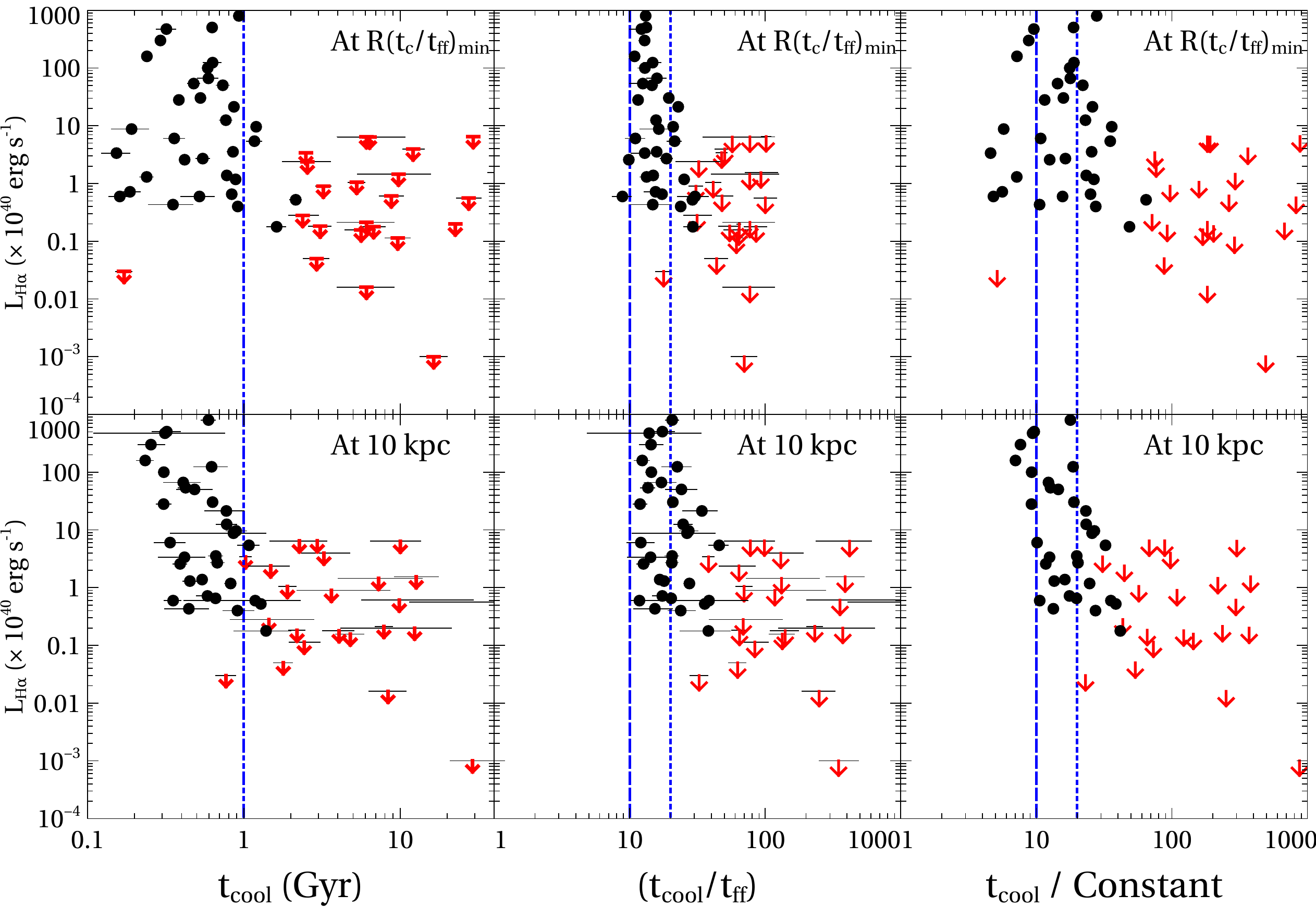}}
  \caption{H$\alpha$ luminosity of the central galaxy as a function of the deprojected cooling time (left), \tctffmin (middle), and the cooling time rescaled to lie in a similar range to \tctffmin\ (right).   The top panels measure each of these quantities at the radius where \tctffmin\ is recorded, \Rtctffmin, whereas the bottom panels show the equivalent quantities taken at a constant radius of 10~kpc.  We note three results.  Firstly the onset of cooling appears no sharper in \tctff\ than in \tc\ alone.  Secondly, no clusters have \tctffmin\ significantly below 10, in tension with predictions.  Thirdly, the range in \tctffmin\ is narrower than the equivalent range in \tc\ when both are measured at the location of the \tctff\ minimum, though the effect is reduced when measured at a single radius.  This narrowing of the range is initially perplexing since there is very little spread in \tff\ (see Section \ref{Section:FFvsC}).  We find the same results if a fixed scale radius of R/R$_{\rm 2500} = 0.02$ is used in place of a fixed physical radius of 10~kpc.}
 \label{Figure:Halpha_Plots}
\end{figure*}

The range in both \tc\ and \tctff\ are shown in Figure \ref{Figure:Halpha_Plots}.  The sharp threshold for the onset of nebular emission is evident. H$\alpha$ luminosity is plotted here against deprojected values of \tc, the minimum of \tctff, and \tc\ rescaled to lie in a similar range to \tctffmin.  These quantities are measured both at \Rtctffmin\ (top panels, Figure \ref{Figure:Halpha_Plots}) and at a radius of 10~kpc (bottom panels, Figure \ref{Figure:Halpha_Plots}).  The H$\alpha$ luminosities were measured heterogeneously and taken from \citet[][]{Crawford99,Cavagnolo09,McDonald10}, and \citet[][]{Rawle12}. Therefore the absolute value of the H$\alpha$ luminosity is uncertain, but the detection of nebular emission indicates cold gas.  A radius of 10~kpc was chosen to ensure most objects are resolved or require only a short extrapolation.  Thermodynamic parameters could instead be presented at a fixed scale radius.  We investigated this by measuring each parameter shown in Figure \ref{Figure:Halpha_Plots} at R/R$_{\rm 2500} = 0.02$.  The results were essentially equivalent to those at 10~kpc.  Our conclusions are therefore not affected by whether we choose to use a fixed physical radius of 10~kpc or fixed scaled radius of 0.02~R$_{\rm 2500}$. Similar distributions in these parameters are found by Pulido et al (Paper II), using CO observations as a cooling indicator in place of H$\alpha$.

Inspection of Figure \ref{Figure:Halpha_Plots} shows that the threshold between the H$\alpha$ emitters and non-emitters is equally sharp for \tc\ alone or \tctff\ (left-hand or middle panels respectively).  However, we have added a variable (\tff) which immediately indicates that the threshold is driven by cooling time.  That only a single cooling cluster lies below \tctffmin$= 10$ is noteworthy, and in this instance it lies below by less than one standard deviation and is thus insignificant.  Therefore one object at most lies below the purported threshold for gas condensation in precipitation models \cite[e.g.][]{McCourt12,Sharma12b,Gaspari13,Prasad15,Choudhury16}.  Similarly, Pulido et al ({\em in prep.}) find that only 1/55 clusters harboring molecular gas reservoirs lies below \tctff$= 10$, and this one again is statistically insignificant.

\begin{figure*}    
  \centering
    \subfigure{\includegraphics[width=14.5cm]{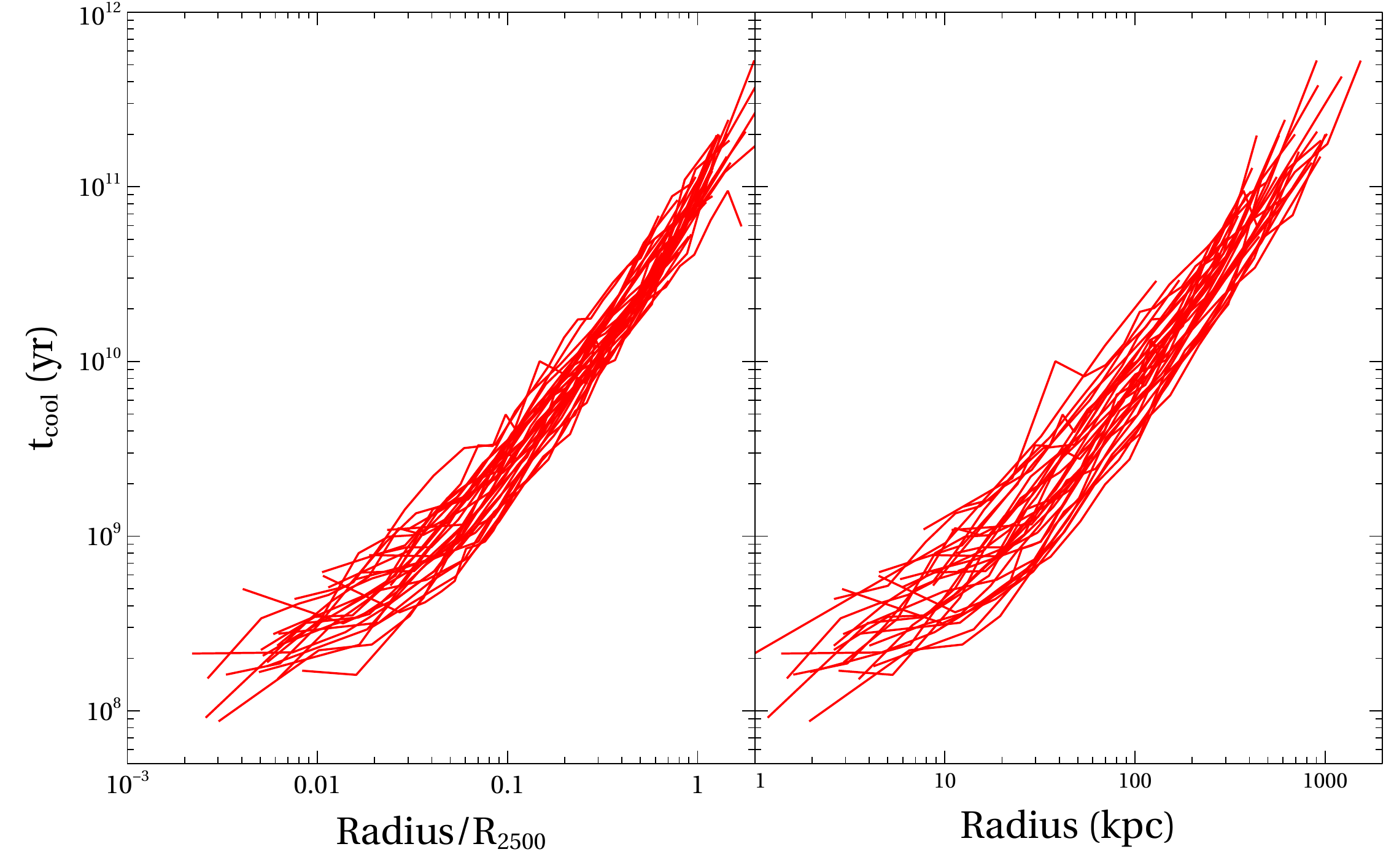}}
  \caption{Cooling time profiles for the line-emitting clusters in our sample.  In the left-hand panel the radius is normalized by the R$_{\rm 2500}$ whereas in the right-hand panel physical radius is plotted.  Rescaling the radius reduces the scatter at both large and small radii.}
 \label{Figure:ScaledCoolingTimeProfiles}
\end{figure*}

The absence of clusters lying below \tctff$= 10$ is a serious problem for precipitation and chaotic cold accretion models. In Figure \ref{Figure:ScaledCoolingTimeProfiles} we plot the cooling time profiles for the H$\alpha$ emitting sources in our sample.  We find a small range in central cooling time, and this range is reduced further when normalized to R$_{\rm 2500}$.  This reflecs the small spread in densities found in Section \ref{Section:DensitySwings}. Furthermore, we find no correlation between the central cooling time and either cavity power or the radio luminosity of the central AGN.  These results, as in Section \ref{Section:DensitySwings}, show that the structure of the ICM is remarkably stable and favor gentle but near continuous AGN feedback.

How rare are departures below \tctff$= 10$ expected to be?  \citet[][]{Li15} simulated a precipitating, self-regulating cool-core cluster that ran for $\sim$6.5 Gyr.  They found that \tctff\ fell below 10, and in the most rapidly cooling phase approached unity, $\sim$25--32\% of the time.  

Observationally, we can take $1 \times 10^{43} {\rm erg s}^{-1}$ as the average cavity power necessary to offset cooling \cite[][]{Rafferty06}.  Converting to an equivalent 1.4~GHz radio-luminosity of $\sim 7 \times 10^{24} {\rm W Hz}^{-1}$ following \citet[][]{Birzan04} and then using the radio luminosity functions presented in \citet[][]{Hogan15a} we find approximately 20--30\% of BCGs host a radio-AGN $\gtrsim 7 \times 10^{24} {\rm W Hz}^{-1}$.  Equating these periods when the AGN launches powerful jets as the period during which feedback is `on', hence when we expect ongoing fuelling of the AGN, then this fraction agrees with the fraction of time with \tctff$<$10 in \citet[][]{Li15}.  If cooling occurs when \tctff$< 10$, as models imply, then  $\sim$25\% of our sample, or roughly 22 clusters from the combined samples here and Paper II, should lie below this threshold, which is overwhelmingly at odds with our measurements.

\subsection{Ratio driven by cooling time} \label{Section:FFvsC}

\begin{figure*}    
  \centering
    \subfigure{\includegraphics[width=8cm]{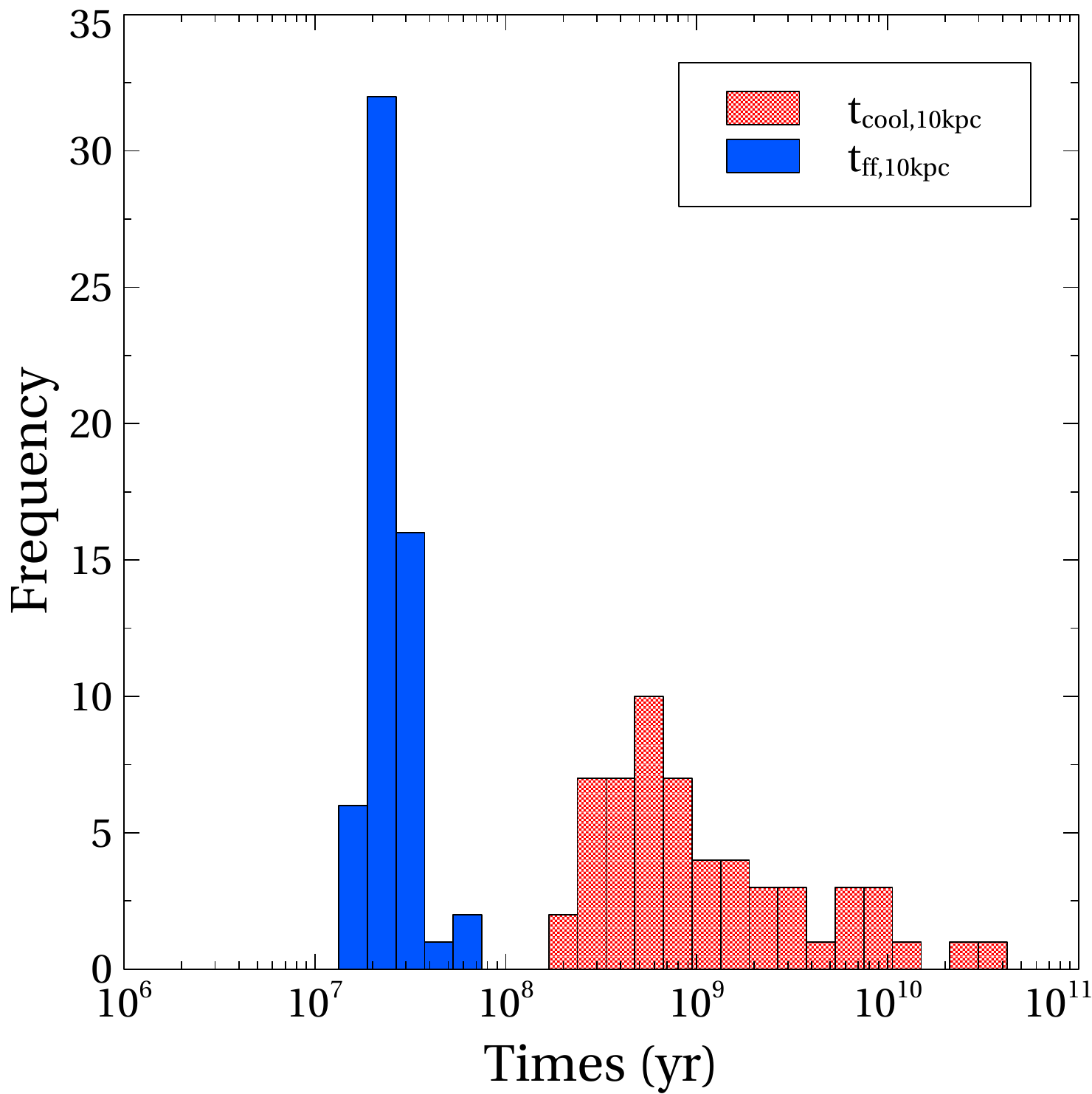}}
    \subfigure{\includegraphics[width=8cm]{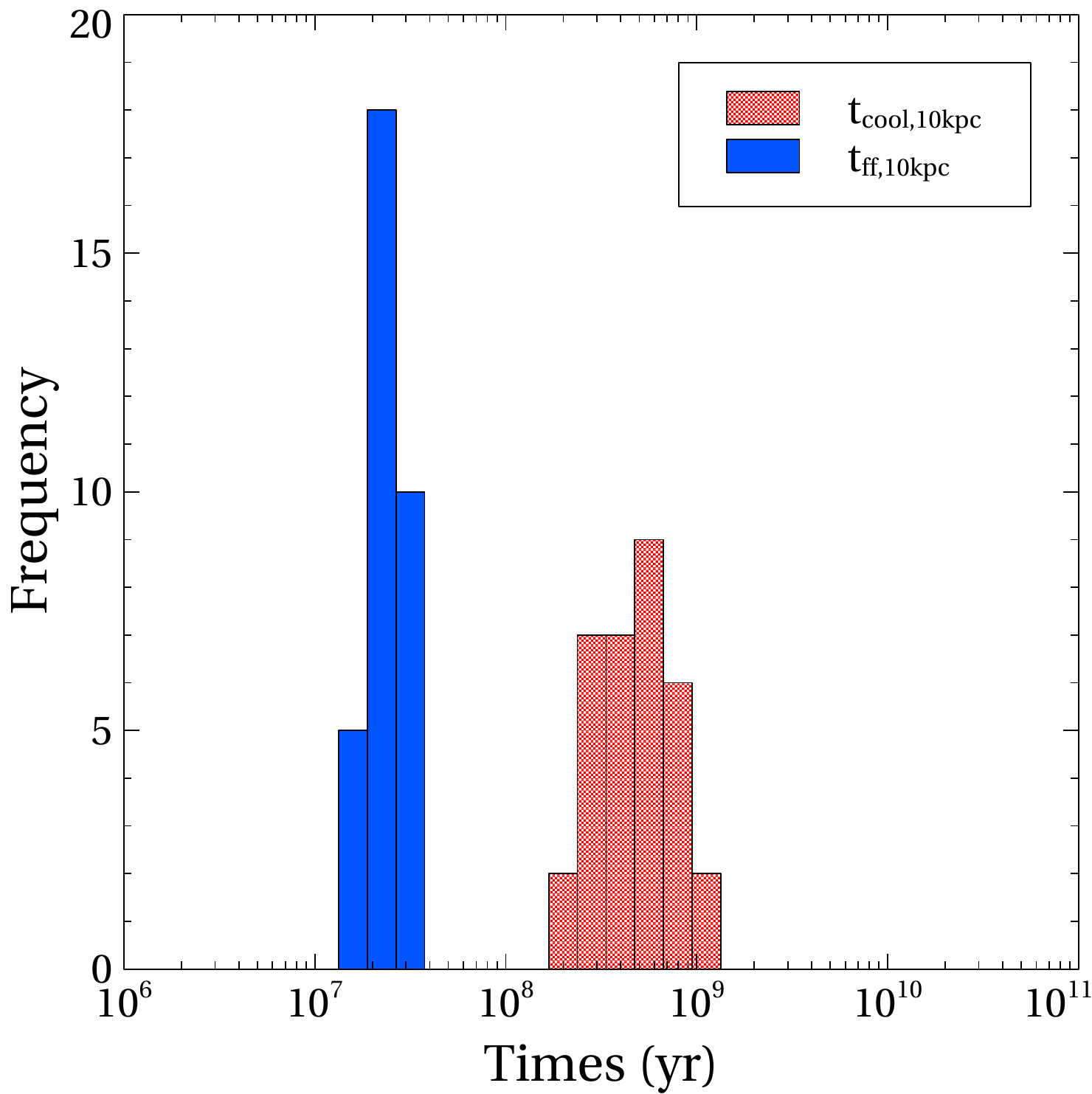}}
  \caption{Spread of cooling and free-fall times at 10~kpc. The left-hand panel shows the full sample whereas the right-hand panel is restricted only to the 33 sources that are known to exhibit H$\alpha$ emission.  Amongst the full sample it is clear that the range in cooling time is much wider than, and therefore dominant over, that of the free-fall time.  When restricted to only the LE clusters the effect is lessened but nevertheless still apparent.}
 \label{Figure:Histogram_DIRECT}
\end{figure*}

\citet[][]{McNamara16} pointed out that the free-fall times found in other studies at the radius of \tctffmin\ span a narrow range.  As a consequence, they showed that the \tctff\ threshold is driven almost entirely by \tc. In Figure \ref{Figure:Histogram_DIRECT} we plot the distributions of free-fall and cooling times of our sample clusters taken at a radius of 10~kpc. Amongst the full sample (left-hand panel) we find a wider spread in cooling time than free-fall time.  The difference in range declines for line-emitting clusters (right-hand panel, Figure \ref{Figure:Histogram_DIRECT}).  The ratio of the standard deviation to the mean is roughly 6 times higher for the distribution of cooling times than the free-fall times.  When only H$\alpha$ emitters are considered this factor falls to 2.5.  Dividing \tc\ by \tff\ is akin to dividing by a constant with a small variance.  Thus the numerator drives the ratio.  It is therefore difficult to understand the role of \tff\ in thermal instability.

In the left-hand panel of Figure \ref{Figure:Timescale_Comparison} we plot both \tc\ and \tff\ at the location of \Rtctffmin, as a function of \tctffmin.  Amongst the full sample we find that both \tc\ and \tff\ are correlated with the ratio, with Kendall’s tau values of 0.71 and 0.56 (both P-value $< 1 \times 10^{-6}$) respectively.  However, the gas in the non cool-core systems is expected to be thermally stable (both \tc\ and \tctff\ lie well above unity) and are largely irrelevant to the argument.  They may also suffer resolution bias in their \tctffmin\ due to the relatively large truncation radii of their cooling profiles. Considering only the H$\alpha$ emitting sub-sample we find \tc\ to be much more dominant in driving the ratio (Kendall’s tau = 0.48, P-value 1.2 × 10−4) than \tff\ (Kendall’s tau = 0.21, P-value 0.1).   The inclusion of \tff\ does not improve the predictive power for the onset of gas cooling above that of the cooling time alone \cite[Section \ref{subsection:Thresholds}, see also][]{McNamara16}. 

\subsubsection{A selection effect can explain the narrow range in \tctffmin} \label{Section:Narrowing}

We show here that the narrow range in minimum \tctff\ is a consequence of the correlation between cooling time, free-fall time, and radius, and the noise imprinted by resolution effects on the measured radius of \tctff\ minimum.  Comparing the spread in L(H$\alpha$) plotted against \tc\ alone (left-hand panel, Figure \ref{Figure:Halpha_Plots}) and \tctff\ (middle panel, Figure \ref{Figure:Halpha_Plots}) may be misleading because of the logarithmic scaling.  We therefore consider instead the standard deviation ($\sigma$) in the \tctffmin\ of the H$\alpha$ emitters  compared to \tc\ normalised by its mean value,  \tc/$<$\tc$>$.

When measured at \Rtctffmin\ we recover a narrower spread in \tctff\ ($\sigma$ = 0.34) than \tc/$<$\tc$>$ ($\sigma$ = 0.65), suggesting that dividing by \tff\ tightens the range.  However, because \tff\ $= R/\sigma$ (see Section \ref{Section:tctff_floor}),  and considering the narrow range of $\sigma$, \Rtctffmin\ is strongly correlated with \tc.  We are thus condemned to measure \tctffmin\ over a narrower range than \tc\ alone at \Rtctffmin.  For example, a measurement of the spread in \tctffmin\ versus the spread in \tc\ at a fixed radius reveals $\sigma$ = 0.43 for \tc/$<$\tc$>$ at 10~kpc, which is much narrower than the spread measured at \Rtctffmin.  Adding to this model-dependent effect is a general bias that the distribution of the minimum of a number of samples of a random variable is narrower than the distribution of the underlying random variable.  Indeed, if we instead take \tctff\ at a fixed physical radius of 10~kpc, the spread in \tctff\ ($\sigma$ = 0.50) is comparable to that in \tc/$<$\tc$>$.

\begin{figure*}    
  \centering
    \subfigure{\includegraphics[width=8cm]{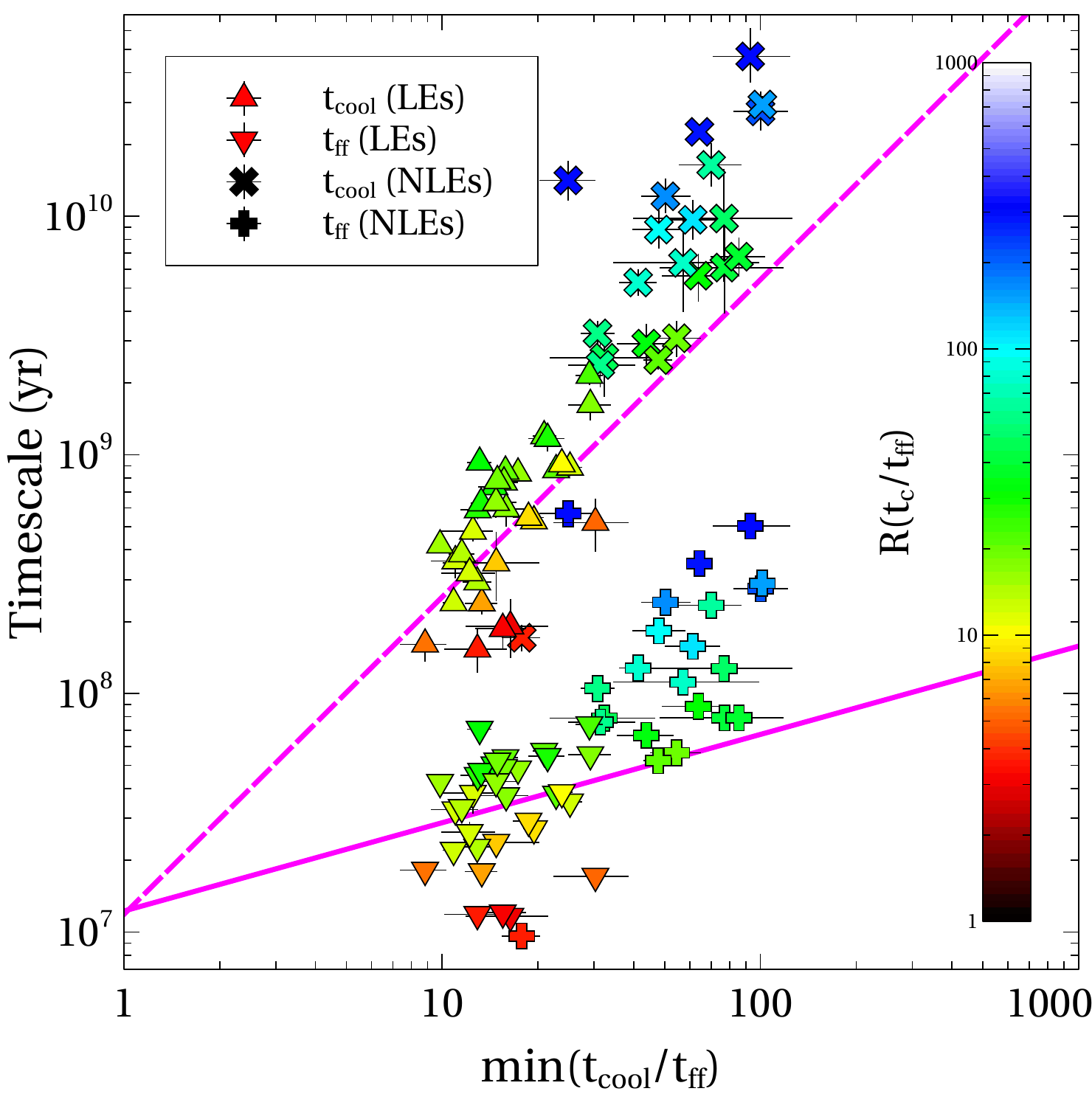}}
    \subfigure{\includegraphics[width=8cm]{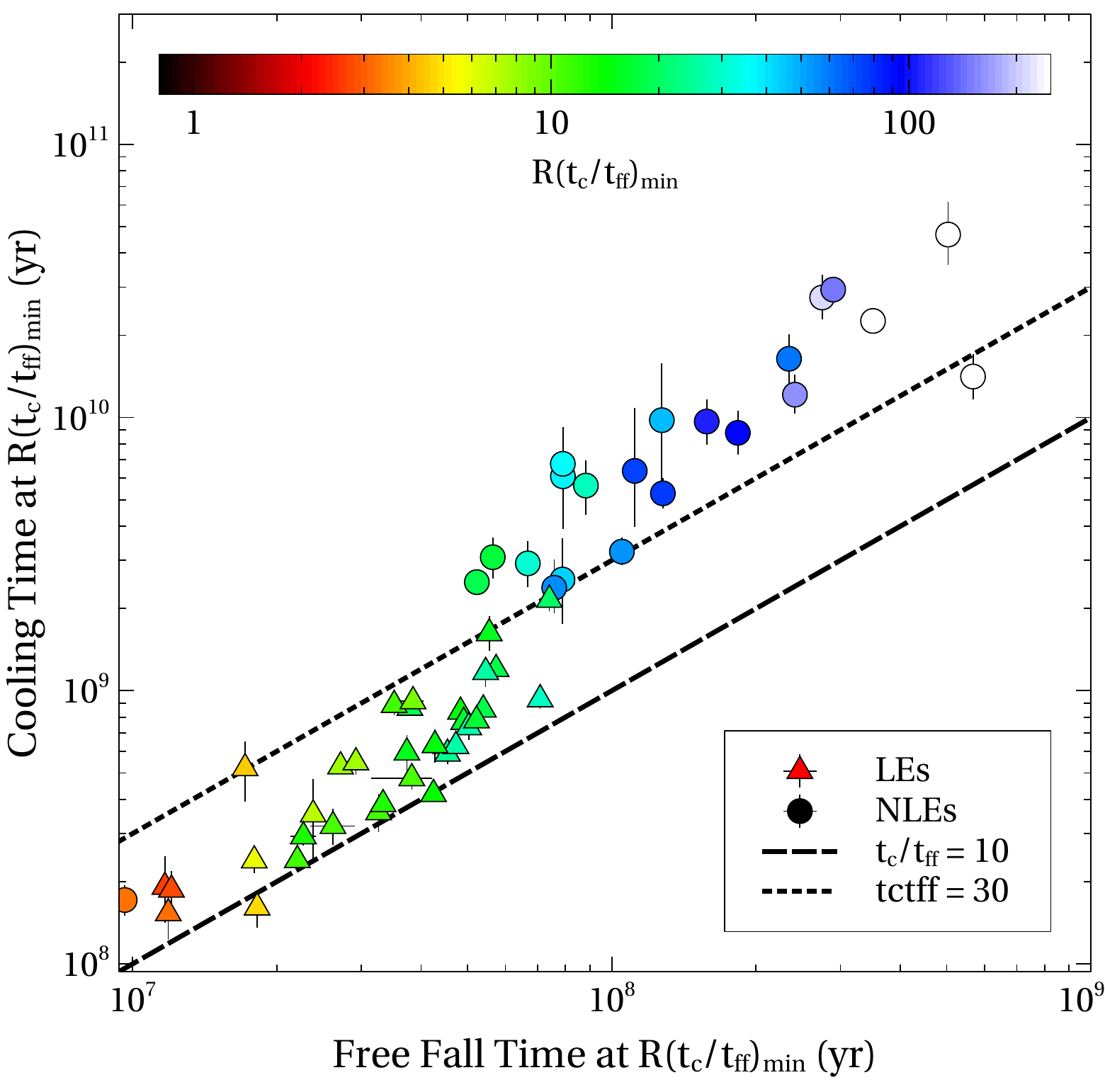}}
  \caption{(Left) The numerator and denominator are plotted against the minimum cooling time to free-fall time ratio.  Non line-emitting (NLE, proxy for non cool-cores) sources appear to show a trend in \tff\ though this is almost certainly bias due to the large truncation radii of their profiles (see Section \ref{Section:Narrowing}).  Solid lines are fits to the LEs (CCs) only.  We find a significant trend only for \tc, showing that the numerator dominates the ratio.  The color gradient shows correlated scatter in \tff\ and \tc\ that can be attributed to their co-dependence on density (see Section \ref{Section:Narrowing}), which itself could explain the narrow range in observed \tctffmin.  (Right) Strong correlation between \tc\ and \tff\ measured at the locations of the \tctff\ minima -- this again could naturally serve to narrow the range of observed \tctffmin, as highlighted by the color gradient with \Rtctffmin.}
 \label{Figure:Timescale_Comparison}
\end{figure*}

This systematic effect is shown clearly in the left-hand panel of Figure \ref{Figure:Timescale_Comparison}.  The points there are color-coded by \Rtctffmin. A matching vertical color-gradient is seen in both \tc\ and \tff\ for any constant value of \tctffmin. This shows that for a given value of \tctffmin\ the ratio \tctff\ is determined by the radius at which it is measured and thus this ratio must lie in a narrow range:  when \tc\ is large, \tff\ is large, and conversely so.  In the right-hand panel of Figure \ref{Figure:Timescale_Comparison} we plot \tc\ vs \tff\ at \Rtctffmin\ (the numerator against the denominator). The sharp lower bound at \tctffmin$=10$ is a consequence of the lowest measured values of the cooling time and the lowest values of the free fall time shown in Figure \ref{Figure:Histogram_DIRECT} differing by a factor of 10.  Adding to this the noise in the estimate of \Rtctffmin\ (Figures \ref{Figure:DeprojectedProfiles_tctctff} \& \ref{Figure:minbin}) and we have the elements of a systematic bias.  Of course, we cannot exclude out of hand that the apparent floor in \tctffmin\ is a natural consequence of feedback \cite[e.g.][]{Voit15b}.  However, a physical floor cannot be disentangled from a systematic bias.  Furthermore, comparisons between \Rtctffmin\ with other thermodynamic properties of interest have failed to reveal correlations (Pulido et al. {\em in prep.}).  The only way to disentangle this bias from a physical correlation would be to identify a sample of galaxies with a broader range of mass (i.e., vary the denominator), which would be difficult.

\section{A Floor Rather than a Minimum in \tctff} \label{Section:tctff_floor}

Here we consider the possibility that the minimum in the \tctff\ profiles may actually be a floor, rather than a clear minimum.  This possibility arises naturally when the mass profile is approximately isothermal within the minimum \tctff\ and when the entropy profile follows a power-law slope of $ K \propto r^{2/3}$ within this region \cite[see Section \ref{Section:EntropyProfiles}, also][]{Panagoulia14a}. We find that both conditions are met in our sample \cite[][and Figure \ref{EntropyProfiles} here]{Panagoulia14a,Hogan17}, and likely in general \cite[e.g.][]{Koopmans09}. 

We observe an inner entropy index $K=Ar^{2/3}$ where A is a constant. From Section \ref{Section:ProjectedProfiles} we have that $n_{\rm e}=(kT/K)^{3/2}$ and $t_{\rm cool}~\propto~P/(n_{\rm e}^{2}\Lambda)$, where cooling function $\Lambda$ depends only on abundance and temperature.  Substituting for pressure and density leads to $t_{\rm cool} \propto K^{3/2}/((kT)^{1/2}\Lambda) \propto (A^{3/2} r)/((kT)^{1/2} \Lambda)$. Using Equation \ref{Equation:tff} and the mass distribution for an isothermal sphere $M=(2\sigma^{2}r)/G$, gives $t_{\rm ff} = r / \sigma $.  Combining these we end up with the radial dependencies cancelling to give ${\rm t}_{\rm cool}/{\rm t}_{\rm ff} \propto (kT \Lambda)^{-1/2}$.  This expression has no dependence on radius, implying that \tctff\ should decline to a constant at a finite radius.  We expect then that the upturns at small radii seen for example in Figures \ref{Figure:ProjectedProfiles_tctctff} \& \ref{Figure:DeprojectedProfiles_tctctff} and in all other studies are produced by density inhomogenities along the line of sight.   

\subsection{Potential Low-altitude Systematic Effects}

\begin{figure}
	\includegraphics[width=\columnwidth]{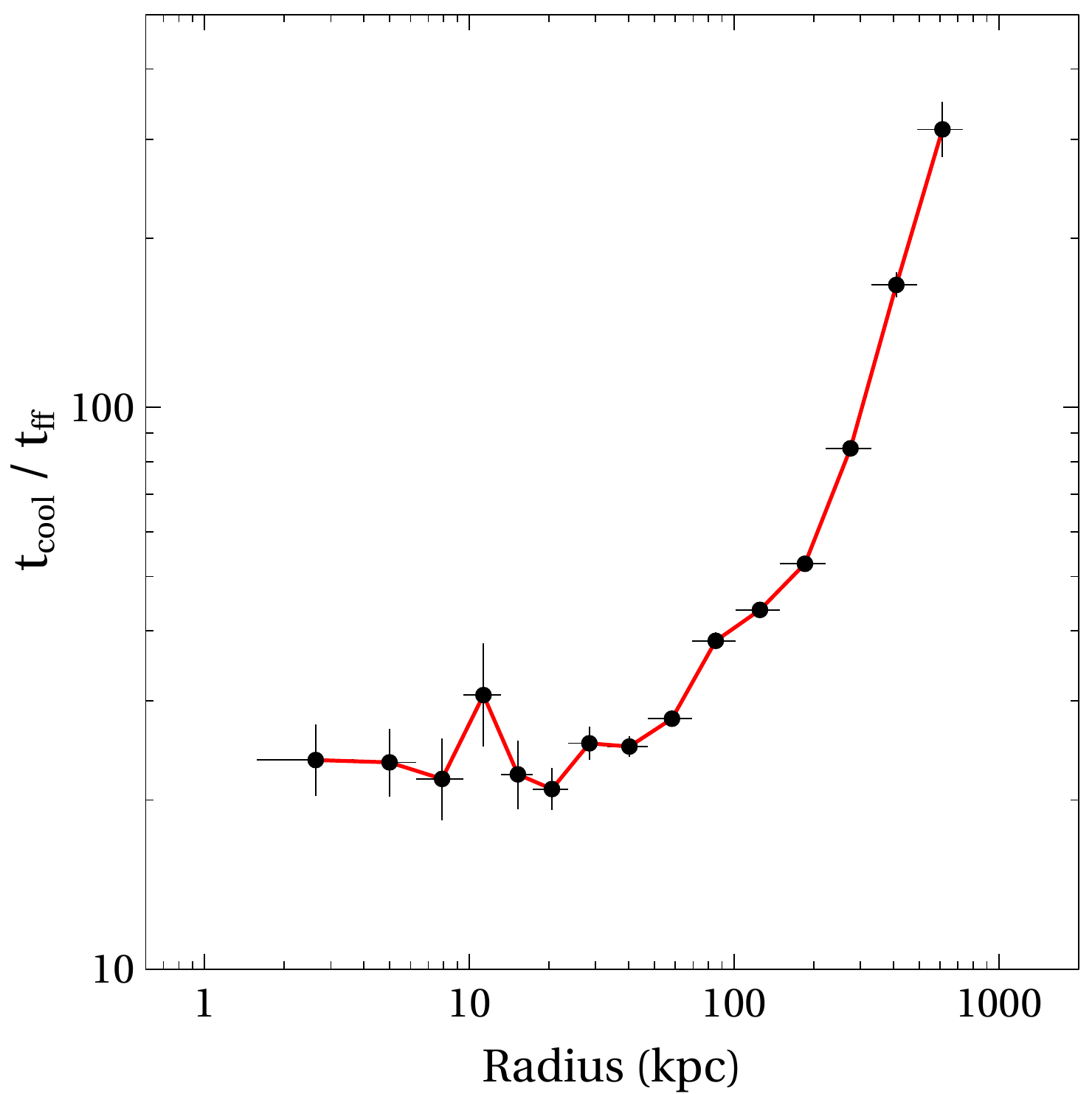}
    \caption{The \tctff profile for the Hydra~A cluster.  This system has the best resolved data near the location of the minimum in this profile.  We do not see a rise to smaller radii, as seen in less well resolved systems.  See discussion in Section \ref{Section:tctff_floor}.}
    \label{Figure:HydraA_tctff}
\end{figure}

Systematic errors induced by limited resolution at small radii in either, or both, the numerator or denominator of \tctff\ could introduce noise and artifically flatten the profile.  For example, over-estimating the inner mass would underestimate \tff.  However, we have attempted to buttress ourselves against this.  Note that our \tctffmin\ are found around 10~kpc.  \citet[][]{Hogan17} derived mass profiles that matched those inferred from stellar velocity dispersions \cite[][]{Fisher95} down to $\sim$1~kpc.  Furthermore, Figure \ref{Figure:Histogram_DIRECT} shows that \tc\ drives the \tctff\ ratio (also see Section \ref{Section:CoolingInstabilities}). We expect most systematics in the inner shape of the profiles concern the measurement of \tc. 

The \tc\ measured in a central annulus (usually with r$_{\rm inner}$=0) may be over-estimated by high temperature gas projected from higher altitudes \cite[][]{Hogan17}.  A more subtle effect concerns the common assumption that \tc\ is approximately constant within a radial shell when creating cooling profiles assuming radial symmetry.  At higher altitudes this assumption is reasonable since the clusters are largely smooth and any structure is averaged over a large volume.  However, almost all cool-core clusters contain cavities in their central atmospheres and the assumption of spherical symmetry breaks down at small radii.  Whilst we attempted to excise highly structured regions from our analysis, it is nearly impossible towards the cluster center.  At these small radii the \tc\ (or \tctff) is likely to vary locally. 

Of the 33 line-emitting clusters in our sample, \tctffmin\ is found in the innermost annulus of 4 (see Figure \ref{Figure:minbin}), which are clearly unresolved.  Seventeen others have only a single spatial bin lying below the position of \Rtctffmin\ and a further 6 have only two bins below the minimum.  The 17 are, likewise, unresolved and the six with two bins are marginally resolved.  Amongst the 6, the value of \tctff\ in the adjacent annulus within \Rtctffmin\ is $< 1\sigma$ greater than the minimum in each case.  To investigate this further, we calculated the equivalent cooling time profiles ${\rm t}_{\rm c,eq}$ that would lead to a flat \tctff\ profile below the \Rtctffmin\ for each of our clusters (i.e. ${\rm t}_{\rm c,eq} = ({t_{\rm c}/t_{\rm ff}})_{\rm min} \times t_{\rm ff}$ ).  We found that the original \tc\ profiles are consistent to within one standard deviation of \tctff\ being flat for 27 of 33 instances.  The inner atmospheres of the remaining six clusters (A478, MS0735+7421, PKS0745-191, Zw2701, Hercules~A, and Zw8276) are inhomogeneous, and thus indeterminate.  It is at least plausible that rises observed below the minima seen here and in other studies are artificial, and that inner \tctff\ profiles are instead flat.  A measurement bias may have previously masked these flat profiles.  Regardless of whether the inner profile is truly flat or not, the minimum value is always lower than any other.  This can create the misleading impression that the inner profile turns upward inside the minimum, particularly when there are few data points at smaller radii.

Amongst this sample the region inside \Rtctffmin\ is best resolved for Hydra~A.  Notably, its \tctff\ profile flattens at its center (Figure \ref{Figure:HydraA_tctff}). This is perhaps surprising considering the expected systematic bias from a heavily structured inner ICM, as is present in this system.  However, Hydra~A shows multiple aligned cavities suggesting that the last few major AGN outbursts are oriented in the same direction \cite[][]{Wise07}, allowing us to isolate undisturbed parts of its atmosphere.  The deep data available permits an accurate measurement of its central \tctff\ profile.  Furthermore, whilst Hydra~A contains a second temperature component in its core, the cooler component contains only a small fraction of the mass \cite[$\sim$1/450 of the hot component,][]{Hogan17}.  Therefore, less uncertainty is expected to be introduced by our single temperature models than for clusters where the cooler component is more noticeable (e.g. A496, ratio $\sim$50 rather than 450). Indeed, amongst the 5 clusters (including Hydra~A) with deep {\it Chandra} data studied in \citet[][]{Hogan17} we found that the deprojected \tctffmin\ were consistent with being flat.  Finally we note that the \tctff\ profile for the nearby cluster M87, which is resolvable to scales $<$1~kpc, also flattens to a floor over 10--15 bins, rather than to a sharp minimum \cite[][]{McNamara16}.

\section{Conclusions} \label{Section:Conclusions}

We have investigated cooling in cluster atmospheres using a sample of 56 clusters.  These were selected as being the most suitable in which to simultaneously study the ICM at low cluster-centric altitudes ($<$10~kpc) and measure accurate total cluster masses to large radii.  Our main findings are as follows:

\begin{itemize}
    \item  \tctff\ gives no more predictive power for the onset of thermally unstable cooling than \tc\ alone.
    \item  Using acceleration profiles that account for the central galaxy's mass, and cooling profiles that use both deprojected density and temperature, we find no cluster atmospheres with a \tctff\ minimum significantly below 10.  Atmospheres with bright nebular emission and star formation lie in the range 10$<$\tctff$<$35, where the upper end of this range corresponds to the \tc\ threshold for thermally unstable gas of 1~Gyr at 10~kpc.
    \item  The absence of clusters with \tctffmin\ $<$ 10 is inconsistent with being a thermodynamic threshold for the onset of cold gas condensation in hot atmospheres.
    \item  The small range of atmospheric gas densities and cooling times at low altitudes indicates AGN heating that is much gentler than predicted by many feedback models.
    \item  The small range in, and measured values of, \tctffmin\ can be attributed to observational biases.  Once the biases are accounted for, the spread in \tctff\ at a fixed altitude is comparable to the spread in \tc\ alone.
    \item  Cool-core entropy profiles are described by a broken power-law: $K \propto r^{0.67}$ between $\sim$1--50~kpc and $K \propto r^{1.1}$ at larger radii.  We find no evidence for flattening below $K \propto r^{0.67}$, or large ($\sim$5--20~kpc) isentropic cores.
    \item  A natural floor in \tctff\ profiles arises from the measured shape of the inner entropy profile and an isothermal mass distribution.  Amongst relaxed/non-merger systems we see no evidence for an upturn in \tctff\ at small radius, and all systems are broadly consistent with a \tctff\ floor.
\end{itemize}

Overall we find that local acceleration in the form of \tff\ provides no additional information concerning gas condensation in galaxy clusters above the cooling time alone. The total cluster mass appears to play a role, and may set the baseline cooling level for the cluster.  This will be investigated in an upcoming paper.  Any weak trends with \tff\ are likely secondary as this parameter effectively traces cluster mass, which may be the true underlying cause. Stimulated feedback via uplift appears to be a promising model for gas condensation but further work is required to test this.

\acknowledgments
Support for this work was provided in part by the National Aeronautics and Space Administration through {\it Chandra} Award Number G05-16134X issued by the {\it Chandra} X-ray Observatory Center. MTH, BRM, FP, ANV, and IB acknowledge support from the Natural Sciences and Engineering Research Council of Canada. HRR acknowledges support from ERC Advanced Grant Feedback 340442. The scientific results reported in this article are based on observations made by the {\it Chandra} X-ray Observatory and has made use of software provided by the {\it Chandra} X-ray Center (CXC) in the application packages CIAO, ChIPS, and Sherpa. The plots in this paper were created using Veusz.

\facilities{Chandra(ACIS)}

\appendix

\section{Tables}

\begin{deluxetable*}{ccccccccc}
\tablecaption{{\it Chandra} data used in our analysis.  Columns are: i) Cluster name, ii) redshift, iii) angular scale on the sky at the given redshifts using standard cosmology, iv) Observation IDs used for the analysis, v) raw combined exposure of the ObsIDs used, vi) useable exposure after data filtering, vii) (fixed) column density used in fitting, viii) RA, ix) DEC.  Sources are presented sorted into regions corresponding to those described in Section \ref{Section:Sample}, then arranged by RA within each region. \label{Table:Observations_Table}}
\tablehead{
\colhead{Cluster} & \colhead{z} & \colhead{Scale}    & \colhead{ObsIDs}  & \multicolumn{2}{c}{Total Exposure}     & \colhead{N$_{\rm H}$}                  &  \multicolumn{2}{c}{Cluster Center}          \\
\colhead{}        & \colhead{}  & \colhead{(kpc/'')} & \colhead{}        & \colhead{Raw}     &  \colhead{Cleaned} & \colhead{(10$^{22}$cm$^{-2}$)}  & \colhead{RA (J2000)} & \colhead{DEC (J2000)} \\
\colhead{}        & \colhead{}  & \colhead{}         & \colhead{}        & \colhead{(ks)}    &  \colhead{(ks)}    & \colhead{}                    & \colhead{}           & \colhead{}          
}
\startdata
 Region A   &        &          &                             &                    &                    &                       &                &               \\
  A85       & 0.0551 & 1.071    & 904,15173,15174,16263,16264 &  195.240           &  193.64            &    0.039              & 00:41:50.476   & -09:18:11.82  \\
  A133      & 0.0566 & 1.098    & 13518,9897,2203             &  154.279           &  141.058           &    0.0153             & 01:02:41.594   & -21:52:53.65  \\
  A401      & 0.0745 & 1.415    & 14024,2309                  &  146.637           &  145.260           &    0.0988             & 02:58:57.862   & +13:34:58.25  \\
  Hydra A   & 0.0550 & 1.069    & 4969,4970                   &  195.734           &  163.79            &    0.043              & 09:18:05.681   & -12:05:43.51  \\
  A1650     & 0.0838 & 1.575    & 4178,5822,6356,6357,        &  190.330           &  167.164           &    0.013              & 12:58:41.485   & -01:45:40.82  \\
            &        &          & 6358,7242,7691              &                    &                    &                       &                &               \\
  A1795     & 0.0625 & 1.204    & 493,3666,5286,5287,5288,       &  666.530        &  625.500           &    0.041              & 13:48:52.521   & +26:35:36.30  \\
            &        &          & 5289,5290,6159,6160,6161,      &                 &                    &                       &                &               \\
            &        &          & 6162,6163,10898,10899,10900,   &                 &                    &                       &                &               \\
            &        &          & 10901,12027,12028,12029,13106, &                 &                    &                       &                &               \\
            &        &          & 13107,13108,13109,13110,13111, &                 &                    &                       &                &               \\
            &        &          & 13112,13113,13412,13413,13414, &                 &                    &                       &                &               \\
            &        &          & 13415,13416,13417,14268,14269, &                 &                    &                       &                &               \\
            &        &          & 14270,14271,14272,14273,14274, &                 &                    &                       &                &               \\
            &        &          & 14275,15485,15486,15487,15488, &                 &                    &                       &                &               \\
            &        &          & 15489,15490                    &                 &                    &                       &                &               \\
  A2029     & 0.0773 & 1.464    & 891,4977,6101               &  107.637           &  103.31            &    0.033              & 15:10:56.077   & +05:44:41.05  \\
  A2142     & 0.0909 & 1.694    & 15186,16564,16565,5005      &  199.709           &  182.308           &    0.0431             & 15:58:19.906   & +27:13:59.36  \\
  Cygnus~A  & 0.0561 & 1.088    & (359),360,(1707),5830,5831, &  243.320           &  228.251           &    0.28               & 19:59:28.316   & +40:44:01.99  \\
            &        &          & 6225,6226,6228,6229,        &                    &                    &                       &                &               \\
            &        &          & 6250,6252                   &                    &                    &                       &                &               \\
  A3667     & 0.0556 & 1.080    & 5751,5752,5753,             &  438.468           &  399.133           &    0.0445             & 20:12:41.710   & -56:51:24.17  \\
            &        &          & 6292,6295,6296              &                    &                    &                       &                &               \\
  A2597     & 0.0852 & 1.598    & 922,7329,6934               &  151.639           &  134.817           &    0.0246             & 23:25:19.720   & -12:07:27.62  \\
  A2626     & 0.0553 & 1.074    & 16136,3192                  &  135.621           &  132.487           &    0.0383             & 23:36:30.432   & +21:08:47.23  \\
  \hline
 Region B   &        &          &                             &                    &                    &                       &                &               \\
  A119      & 0.0442 &   0.870  & 4180,7918                   &  56.971            &   55.947           &    0.0352             & 00:56:16.088   & -01:15:20.37  \\
  A160      & 0.044  &   0.866  & 3219                        &  58.491            &   51.326           &    0.0406             & 01:12:59.749   & +15:29:28.53  \\
  A3112     & 0.0761 &   1.443  & 13135,2516                  &  59.164            &   51.832           &    0.0133             & 03:17:57.654   & -44:14:17.97  \\
  A478      & 0.0881 &   1.647  & 1669,6102                   &  52.390            &   46.763           &    0.281              & 04:13:25.291   & +10:27:55.15  \\
  A3376     & 0.0456 &   0.896  & 3202,3450                   &  64.115            &   62.070           &    0.0498             & 06:02:10.700   & -39:57:37.05  \\
  A1644     & 0.0471 &   0.924  & 2206,7922                   &  70.199            &   70.035           &    0.0413             & 12:57:11.564   & -17:24:34.76  \\
  Zw8276    & 0.075  &   1.424  & 11708,8267                  &  53.474            &   53.474           &    0.0383             & 17:44:14.453   & +32:59:29.41  \\
  A2319     & 0.0557 &   1.082  & 3231,15187                  &  89.600            &   86.793           &    0.0810             & 19:21:10.110   & +43:56:44.20  \\
  AS1101    & 0.058  &   1.123  & 11758                       &  97.735            &   92.573           &    0.039              & 23:13:58.693   & -42:43:38.58  \\
  A2589     & 0.0407 &   0.805  & 6948,7190,7340              &  78.666            &   78.439           &    0.0316             & 23:23:57.356   & +16:46:38.55  \\
  A4059     & 0.0475 &   0.931  & 5785                        &  92.121            &   87.938           &    0.012              & 23:57:00.473   & -34:45:33.04  \\
  \hline
 Region C   &        &          &                             &                    &                    &                       &                &               \\
PKS0745-191 & 0.1028 &   1.890  & 12881,2427,6103,7694        &  151.189           &   148.629          &     0.415             & 07:47:31.291   & -19:17:40.02  \\
  A1413     & 0.1427 &   2.508  & 1661,(5002),5003            &  121.456           &   106.103          &     0.0183            & 11:55:17.991   & +23:24:19.82  \\
  A2034     & 0.111  &   2.022  & 12885,12886,13192,2204      &  250.951           &   226.072          &     0.0154            & 15:10:11.556   & +33:30:40.53  \\
Hercules~A  & 0.154  &   2.672  & 1625,5796,6257              &  111.86            &   108.792          &     0.06              & 16:51:08.203   & +04:59:32.51  \\
  \hline
 Region D   &        &          &                             &                    &                    &                       &                &               \\
RXJ0821+07  & 0.110  &  2.006   & 17194,17563                 &  66.559            &   63.488           &    0.0195             & 08:21:02.242   & +07:51:48.99  \\
  A1201     & 0.1688 &   2.881  & 4216,9616                   &  87.061            &   63.581           &    0.0157             & 11:12:54.536   & +13:26:07.75  \\
  A2069     & 0.1138 &   2.066  & 4965                        &  55.417            &   46.198           &    0.0192             & 15:24:07.476   & +29:53:17.42  \\
  A2204     & 0.1522 &  2.646   & 7940,499                    &  87.210            &   80.563           &    0.061              & 16:32:46.887   & +05:34:31.42  \\
  A2244     & 0.0980 &   1.812  & 4179                        &  56.965            &   53.894           &    0.0188             & 17:02:42.357   & +34:03:36.51  \\
  \hline
 Region E   &        &          &                             &                    &                    &                       &                &               \\
  A399      & 0.0716 &   1.365  & 3230                        &  48.631            &   46.328           &    0.106              & 02:57:53.124   & +13:01:51.09  \\
  A576      & 0.0385 &   0.763  & 3289                        &  38.592            &   27.737           &    0.055              & 07:21:30.162   & +55:45:41.71  \\
  A744      & 0.0729 &   1.387  & 6947                        &  39.519            &   34.596           &    0.0343             & 09:07:20.481   & +16:39:04.56  \\
  NGC5098   & 0.0394 &   0.780  & 6941                        &  38.623            &   38.623           &    0.0131             & 13:20:14.728   & +33:08:36.15  \\   
  A3571     & 0.0391 &   0.774  & 4203                        &  33.987            &   15.680           &    0.0425             & 13:47:28.599   & -32:51:54.71  \\
  A1991     & 0.0587 &   1.136  & 3193                        &  38.305            &   34.526           &    0.0234             & 14:54:31.554   & +18:38:38.29  \\
  A2107     & 0.042  &   0.829  & 4960                        &  35.573            &   34.805           &    0.0445             & 15:39:39.043   & +21:46:58.55  \\
  \hline                                                                                                                      
 Region F   &        &          &                             &                    &                    &                       &                &               \\
MS0735+7421 & 0.216  &   3.503  & 10468,10469,10470,10471     &  476.700           &   446.943          &    0.0328             & 07:41:44.205   & +74:14:38.31  \\
            &        &          & 10822,10918,10922           &                    &                    &                       &                &               \\
 A665       & 0.1824 &   3.067  & 12286,13201,3586            &  125.528           &   108.598          &    0.0433             & 08:30:58.622   & +65:50:24.49  \\
 4C+55.16   & 0.242  &   3.817  &  1645,4940                  &  106.030           &   70.488           &    0.0449             & 08:34:54.845   & +30:20:59.43  \\
 Zw2701     & 0.215  &   3.490  & 12903,3195                  &  122.685           &   117.568          &    0.00751            & 09:52:49.161   & +51:53:05.58  \\
 A1689      & 0.1832 &   3.078  & 1663,5004,6930,7289         &  181.857           &   165.773          &    0.0183             & 13:11:29.512   & -01:20:28.03  \\
 A1758      & 0.279  &   4.233  & 13997,15538,15540           &  147.696           &   129.722          &    0.0103             & 13:32:48.548   & +50:32:32.71  \\
MACS1347-11 & 0.451  &   5.767  & 13516,13999,14407,3592,     &  233.800           &   204.415          &    0.046              & 13:47:30.582   & -11:45:09.21  \\
            &        &          & 506,507                     &                    &                    &                       &                &               \\
 A1835      & 0.2532 &   3.947  & 6880,6881,7370,495,496      &  252.740           &   204.217          &    0.020              & 14:01:02.080   & +02:52:42.99  \\
MACS1423+24 & 0.543  &   6.370  & 1657,4195                   &  134.095           &   121.766          &    0.022              & 14:23:47.870   & +24:04:42.50  \\
 Zw7160     & 0.2578 &   3.999  &  4192,543,7709              &  108.804           &   92.179           &    0.0318             & 14:57:15.104   & +22:20:33.89  \\
MACS1532+30 & 0.343  &   4.875  & 14009,1649,1665             &  108.198           &   102.191          &    0.023              & 15:32:53.747   & +30:20:59.43  \\
 A2219      & 0.2248 &   3.611  & 13988,14355,14356,          &  189.741           &   173.443          &    0.0176             & 16:40:19.822   & +46:42:41.19  \\
            &        &          & 14431,14451,896             &                    &                    &                       &                &               \\
 A2390      & 0.228  &   3.650  & (500),(501),4193            &  113.94            &   71.647           &    0.079              & 21:53:36.792   & +17:41:44.25  \\
  \hline
 Region G   &        &          &                             &                    &                    &                       &                &               \\
RXJ0338+09  & 0.0349 &   0.695  & 7939,919,9792               &  103.012           &  91.748            &    0.176              & 03:38:40.597   & +09:58:12.54  \\
    A496    & 0.0329 & 0.656    & 931,3361$^{+}$,4976          &  104.002           &  62.750            &    0.040              & 04:33:37.932   & -13:15:40.59  \\
    A2052   & 0.0355 &   0.706  & 890,5807,10477,10478,       &  644.990           &  640.429           &    0.027              & 15:16:44.484   & +07:01:17.86   \\
            &        &          & 10479,10480,10879,10914     &                    &                    &                       &                &               \\
            &        &          & ,10915,10916,10917          &                    &                    &                       &                &               \\
    A2199   & 0.0302 & 0.605    & 10748,10803,10804,10805     & 119.870            &  119.610           &    0.039              & 16:28:38.245   & +39:33:04.21  \\
            &        &          & ,10804,10805                &                    &                    &                       &                &               \\
 IC1262     & 0.0331 &   0.660  & 2018,6949,7321,7322         &  144.430           &  130.164           &    0.0178             & 17:33:01.973   & +43:45:35.13  \\
\enddata
\end{deluxetable*}

\begin{deluxetable*}{ccccccccc}
\tablecaption{Details of the \textsc{isonfwmass} profile fits.  Columns are: i) Cluster name, ii) Line-emitting (LE) or non line-emitting (NLE) BCG, indicative of cool-core, iii) equivalent stellar velocity dispersion, iv) isothermal potential = $\mu$m$_{H}$$\sigma^{2}$ where m$_{H}$ is the mass of the hydrogen atom and the mean atomic weight $\mu$=0.59, v)  NFW scale radius, vi) NFW potential = 4$\pi$G$\rho_{0}$R$_{s}^{2}$~$\mu$M$_{H}$ in units of keV, vii) R$_{2500}$, viii) M$_{2500}$.  The reported $\rho_{0,ISO}$ values correspond to the $\sigma_{*}$ values and were kept fixed in the fitting to account for the anchored stellar mass component. Sources are ordered as in Table \ref{Table:Observations_Table}.  \label{Table:ISONFW_fits}}
\tablehead{
\colhead{Cluster} & \colhead{Lines?} & \colhead{$\sigma_{*}$} & \colhead{$\rho_{0,ISO}$} & \colhead{Beta} & \colhead{R$_{s,NFW}$} & \colhead{$\rho_{0,NFW}$} & \colhead{R$_{2500}$} & \colhead{M$_{2500}$}                     \\
\colhead{}        & \colhead{} & \colhead{(km/s)}      & \colhead{(keV)}         & \colhead{}     & \colhead{(arcmin)}  &  \colhead{(keV)}        & \colhead{(kpc)}    & \colhead{($\times$10$^{14}$~M$_{\odot}$)} \\
\colhead{}        & \colhead{} & \colhead{}            & \colhead{}              & \colhead{}     & \colhead{}          &  \colhead{}             & \colhead{}         & \colhead{}           
}
\startdata
    Region A &                &               &        &                      &                       &        &                     \\[5pt] 
    A85     &  LE & 270.4$\pm$6.4   &   0.450       &   -    & 7.37$^{+0.46}_{-0.21}$  &  49.24$^{+1.64}_{-0.92}$ & 516.7  & 2.07$^{+0.04}_{-0.03}$ \\[5pt] 
    A133    &  LE & 249.1$\pm$7.6   &   0.382       & 0.60   & 13.86$^{+1.18}_{-0.74}$ &  64.90$^{+3.80}_{-1.71}$ & 519.6  & 2.10$^{+0.08}_{-0.09}$ \\[5pt] 
    A401    & NLE & 280.3$\pm$8.8   &   0.486       &  0.50  & 7.53$^{+0.30}_{-0.16}$  &  73.18$^{+1.30}_{-2.96}$ & 565.1 & 2.82$^{+0.10}_{-0.10}$   \\[5pt] 
    Hydra A &  LE & 236.6$\pm$8.4   &   0.344       &   -    & 5.85$^{+0.53}_{-0.49}$  &  32.29$^{+1.52}_{-1.55}$ & 423.6  & 1.14$^{+0.03}_{-0.04}$ \\[5pt] 
    A1650   & NLE & 236.1$\pm$12.4  &   0.343       &   -   & 3.11$^{+0.11}_{-0.15}$  &  46.64$^{+0.80}_{-0.78}$  & 506.2  & 2.00$^{+0.07}_{-0.06}$ \\[5pt] 
    A1795   &  LE & 221.1$\pm$6.5   &   0.302       &   -    & 7.45$^{+0.25}_{-0.25}$  &  57.29$^{+1.18}_{-1.25}$ & 539.3 & 2.37$^{+0.03}_{-0.04}$  \\[5pt] 
    A2029   & NLE & 335.9$\pm$10.0  &   0.694       &   -    & 6.79$^{+0.49}_{-0.46}$  &  88.68$^{+4.84}_{-4.10}$ & 686.1  & 4.94$^{+0.17}_{-0.19}$ \\[5pt] 
   A2142    & NLE & 241.2$\pm$11.3  &   0.360       &   -   & 14.67$^{+1.58}_{-0.16}$ & 156.47$^{+13.92}_{-2.00}$ &  753.2 &  6.63$^{+0.46}_{-0.39}$ \\[5pt]
   Cygnus A &  LE & 268.5$\pm$7.5   &   0.446       &   -   & 2.43$^{+0.22}_{-0.18}$   &  48.44$^{+1.26}_{-0.82}$ &  525.8 & 2.18$^{+0.06}_{-0.06}$  \\[5pt] 
    A3667   & NLE & 262.3$\pm$7.6   &   0.426       &   -    & 17.36$^{+1.19}_{-0.59} $ & 50.46$^{+2.25}_{-0.52}$ & 418.1  &  1.10$^{+0.14}_{-0.13}$ \\[5pt] 
    A2597   &  LE & 217.7$\pm$10.4  &   0.293       &   -    & 2.86$^{+0.15}_{-0.08}$  &  36.63$^{+1.41}_{-0.59}$ & 452.3 & 1.43$^{+0.03}_{-0.03}$  \\[5pt] 
    A2626   &  LE & 243.3$\pm$7.3   &   0.366       &   -    & 2.56$^{+0.08}_{-0.13}$  &  18.51$^{+0.25}_{-0.38}$ & 344.5 &  0.61$^{+0.14}_{-0.14}$ \\[5pt]
  \hline      
    Region B &                &               &        &                      &                      &        &                      \\[5pt]
    A119     & NLE & 237.9$\pm$5.4  &   0.350       &   -   & 4.49$^{+3.53}_{-3.83}$  &  17.58$^{+3.16}_{-14.98}$ &  333.1 & 0.58$^{+0.08}_{-0.08}$  \\[5pt] 
    A160     & NLE & 207.3$\pm$7.3  &   0.266       &   -   &  2.07$^{+0.35}_{-0.58}$ &  11.93$^{+0.41}_{-0.46}$  &  280.1 & 0.33$^{+0.02}_{-0.02}$  \\[5pt] 
    A3112    &  LE & 266.2$\pm$8.5  &   0.438       &   -    & 7.50$^{+0.63}_{-0.84}$  &  66.70$^{+4.39}_{-61.48}$   &  570.5 & 2.84$^{+0.14}_{-0.15}$ \\[5pt] 
    A478     &  LE & 271.1$\pm$7.0  &   0.455       &   -   &  7.10$^{+0.17}_{-0.24}$ &  86.66$^{+1.75}_{-1.73}$  & 647.0  & 4.19$^{+0.59}_{-0.66}$ \\[5pt] 
    A3376*   & NLE & 198.5$\pm$7.0  &  0.01*(0.244) &   -   & 14.99$^{+7.32}_{-1.10}$ &  24.36$^{+7.02}_{-2.10}$ &  263.1 &  0.27$^{+0.03}_{-0.03}$ \\[5pt] 
   A1644     &  LE & 248.5$\pm$6.6  &   0.382       &   -   & 6.11$^{+1.98}_{-1.83}$  & 10.67$^{+2.64}_{-2.32}$  &  258.3 &  0.26$^{+0.03}_{-0.03}$ \\[5pt] 
    Zw8276   &  LE & 218.0$\pm$7.1  &   0.294       &   -   & 3.35$^{+0.33}_{-0.33}$  &  36.78$^{+1.97}_{-1.23}$ &  454.6 & 1.43$^{+0.05}_{-0.05}$  \\[5pt] 
    A2319    & NLE & 249.1$\pm$6.5  &   0.384       &   -   &  9.91$^{+0.91}_{-1.50}$ &  81.22$^{+4.55}_{-8.35}$ &  642.1 & 3.97$^{+0.24}_{-0.23}$  \\[5pt] 
   AS1101    &  LE & 219.1$\pm$7.5  &   0.297       &   -   & 2.77$^{+0.12}_{-0.08}$  &  21.82$^{+0.46}_{-0.22}$ &  364.6 &  0.73$^{+0.01}_{-0.01}$ \\[5pt] 
    A2589    & NLE & 220.2$\pm$6.3  &   0.300       &   -   & 6.38$^{+0.44}_{-0.68}$  &  27.47$^{+1.15}_{-1.61}$  &  398.7 & 0.94$^{+0.03}_{-0.03}$ \\[5pt] 
    A4059    &  LE & 244.3$\pm$5.7  &   0.369       &   -   &  4.33$^{+0.43}_{-0.61}$ &  29.98$^{+1.13}_{-1.89}$ &  425.6 &  1.15$^{+0.05}_{-0.04}$ \\[5pt] 
  \hline      
    Region C &                 &              &       &                      &                       &        &                      \\[5pt]
 PKS0745-191 &  LE & 289.8$\pm$14.3  &  0.519       &   -   & 4.32$^{+0.23}_{-0.38}$  &  75.38$^{+1.71}_{-3.65}$ &  629.1 & 3.91$^{+0.10}_{-0.10}$        \\[5pt] 
    A1413+   & NLE & 363.7$\pm$12.4  &  0.818       &   -   & 5.77$^{+0.15}_{-0.03}$  &  98.51$^{+0.66}_{-2.77}$ &  660.0 & 4.70$^{+0.15}_{-0.15}$  \\[5pt] 
    A2034*   & NLE & 276.6$\pm$10.8  & 0.01*(0.473) &   -   & 31.69$^{+14.31}_{-1.87}$ & 283.09$^{+122.52}_{-17.31}$ & 699.4 & 5.41$^{+0.59}_{-0.54}$ \\[5pt]
 Hercules A  &  LE & 284.7$\pm$13.9  &  0.501       &   -   & 0.91$^{+0.01}_{-0.10}$  &  27.48$^{+0.58}_{-0.92}$  &  395.7 & 1.02$^{+0.03}_{-0.04}$ \\[5pt] 
  \hline      
    Region D &                 &              &       &                      &                      &        &                      \\[5pt]
 RXJ0821+07  &  LE & 246.7$\pm$8.9   &   0.376      &   -   & 1.56$^{+0.53}_{-0.48}$  &  21.48$^{+2.45}_{-1.99}$ & 357.7  & 0.72$^{+0.06}_{-0.07}$  \\[5pt] 
    A1201    & NLE & 338.1$\pm$12.8  &   0.707      &   -   & 10.35$^{+2.33}_{-2.35}$ & 81.06$^{+15.51}_{-16.69}$ &  446.6 & 1.50$^{+0.14}_{-0.13}$  \\[5pt] 
    A2069*   & NLE & 262.8$\pm$10.0  & 0.01*(0.427) &   -   & 2.91$^{+0.89}_{-0.68}$  &  19.12$^{+1.66}_{-2.22}$  & 323.7  & 0.54$^{+0.04}_{-0.05}$  \\[5pt] 
    A2204    &  LE & 343.3$\pm$13.0  &   0.729      &   -   & 1.25$^{+0.14}_{-0.19}$  &  78.54$^{+1.80}_{-2.34}$ & 639.7  & 4.32$^{+0.16}_{-0.15}$   \\[5pt] 
    A2244    & NLE & 288.1$\pm$8.5   &   0.513      &   -   & 3.58$^{+0.43}_{-0.36}$  &  45.77$^{+2.93}_{-2.26}$  &  501.0  & 1.96$^{+0.07}_{-0.07}$ \\[5pt]
  \hline      
    Region E &                &               &        &                      &                 &        &                      \\[5pt]
    A399     & NLE & 269.0$\pm$8.9  &     0.448     &   -   & 3.91$^{+0.43}_{-0.98}$  &  39.71$^{+2.59}_{-4.44}$  &  478.2 & 1.66$^{+0.10}_{-0.10}$  \\[5pt] 
    A576     & NLE & 224.9$\pm$4.5  &     0.313     &   -   & 15.62$^{+4.41}_{-4.45}$ &  42.05$^{+9.67}_{-9.55}$  &  431.3 & 1.18$^{+0.15}_{-0.13}$ \\[5pt] 
    A744     & NLE & 247.3$\pm$6.6  &     0.378     &   -   & 1.12$^{+0.35}_{-0.19}$  &  12.03$^{+1.49}_{-0.90}$  &  285.3 & 0.35$^{+0.04}_{-0.04}$ \\[5pt] 
    NGC5098  &  LE & 186.0$\pm$4.8  &     0.214     &   -   & 0.96$^{0.61+}_{-0.22}$  &   4.47$^{+0.51}_{-0.53}$  &  183.9 & 0.09$^{+0.01}_{-0.01}$ \\[5pt] 
    A3571    & NLE & 253.1$\pm$4.9  &     0.396     &   -   & 9.80$^{+1.88}_{-1.54}$  &  56.54$^{+6.90}_{-5.57}$  &  559.7 & 2.59$^{+0.21}_{-0.20}$  \\[5pt] 
    A1991    &  LE & 221.7$\pm$8.0  &     0.304     &  0.44 & 1.64$^{+0.12}_{-0.07}$  &  17.89$^{+0.12}_{-0.08}$ &  331.9 & 0.55$^{+0.01}_{-0.01}$  \\[5pt] 
    A2107    & NLE & 246.7$\pm$5.1  &     0.377     &   -   & 4.00$^{+0.66}_{-0.65}$  &  27.72$^{+1.95}_{-1.82}$  &  413.9 & 1.05$^{+0.05}_{-0.06}$ \\[5pt] 
  \hline      
    Region F &                &               &       &                      &                      &        &                      \\[5pt]
MS0735+7421  &  LE & 314.5$\pm$17.4 &     0.612     &   -   & 7.25$^{+0.06}_{-0.30}$  &  96.28$^{+0.57}_{-2.65}$  &  507.7  & 2.31$^{+0.08}_{-0.08}$ \\[5pt] 
   A665      & NLE & 362.8$\pm$7.0$^{1}$ &  0.810   &   -   & 9.47$^{+3.11}_{-1.95}$  &  95.62$^{+23.02}_{-15.59}$ &  501.9 & 2.15$^{+0.23}_{-0.25}$ \\[5pt] 
  4C+55.16   &  LE & 274.0$\pm$24.1 &     0.464     &   -   & 2.06$^{+0.72}_{-0.39}$  &  47.52$^{+9.84}_{-6.27}$  &  455.2 & 1.71$^{+0.19}_{-0.17}$ \\[5pt] 
   Zw2701    &  LE & 340.8$\pm$17.3 &     0.718     &   -   & 1.37$^{+0.23}_{-0.22}$  &  41.21$^{+2.37}_{-2.58}$  &  466.6  & 1.79$^{+0.08}_{-0.09}$ \\[5pt] 
   A1689     & NLE & 355.3$\pm$17.7 &     0.781     &   -   & 2.44$^{+0.61}_{-0.06}$  & 102.73$^{+7.95}_{-1.03}$  & 718.2  & 6.31$^{+0.18}_{-0.34}$  \\[5pt] 
   A1758*    & NLE & 376.5$\pm$21.0 &   0.01(0.877) &  0.81 & 1.22$^{+0.53}_{-0.32}$  &  17.56$^{+4.83}_{-3.39}$  &  251.7 & 0.30$^{+0.04}_{-0.04}$  \\[5pt] 
MACS1347-11  &  LE & 362.8$\pm$7.0$^{1}$ &  0.810   &   -   & 1.06$^{+0.20}_{-0.08}$  & 163.23$^{+12.17}_{-5.25}$ &  776.3 & 10.77$^{+0.07}_{-0.07}$ \\[5pt] 
   A1835     &  LE & 485.6$\pm$24.2 &     1.458     &   -   & 5.46$^{+0.51}_{-0.78}$  & 143.91$^{+11.17}_{-16.71}$ & 711.3 & 6.61$^{+0.37}_{-0.38}$  \\[5pt] 
MACS1423+24  &  LE & 362.8$\pm$7.0$^{1}$ &  0.810   &   -   & 0.96$^{+0.33}_{-0.10}$  &  74.93$^{+11.86}_{-5.67}$ &  501.8 & 3.24$^{+0.03}_{-0.03}$  \\[5pt] 
  Zw7160     &  LE & 428.1$\pm$20.5 &     1.134     &   -   & 1.44$^{+0.23}_{-0.15}$  &  47.28$^{+3.49}_{-2.82}$  &  497.8 & 2.28$^{+0.11}_{-0.11}$ \\[5pt] 
MACS1532+30  &  LE & 362.8$\pm$7.0$^{1}$ &  0.810   &  -    & 1.90$^{+0.38}_{-0.24}$  &  81.95$^{+10.80}_{-5.82}$ &  570.1 & 3.76$^{+0.32}_{-0.32}$ \\[5pt] 
   A2219     & NLE & 342.5$\pm$21.9 &     0.726     &   -   & 4.52$^{+0.51}_{-0.48}$  & 120.63$^{+8.78}_{-8.78}$  &  678.8 & 5.57$^{+0.23}_{-0.23}$ \\[5pt] 
   A2390     &  LE & 348.2$\pm$22.5 &     0.750     &   -   & 4.68$^{+0.70}_{-0.74}$  & 118.75$^{+13.15}_{-13.09}$ & 664.3 & 5.24$^{+0.37}_{-0.34}$  \\[5pt] 
  \hline      
    Region G &                &               &        &                      &                      &        &                      \\[5pt]
RXJ0338+09   &  LE & 215.4$\pm$4.8   &   0.287       &   -   & 4.98$^{+0.38}_{-0.17}$   &  29.12$^{+1.08}_{-0.36}$  &  418.9  & 1.08$^{+0.03}_{-0.04}$  \\[5pt]
    A496     &  LE & 228.1$\pm$4.6   &   0.320       &   -    & 14.00$^{+2.88}_{-2.09}$ &  45.65$^{+5.93}_{-3.68}$ & 482.5  & 1.65$^{+0.11}_{-0.11}$ \\[5pt]  
    A2052    &  LE & 221.1$\pm$5.4   &   0.302       &  0.62  & 5.55$^{+0.17}_{-0.27}$       &  25.65$^{+0.15}_{-0.06}$    &  394.2 &  0.90$^{+0.09}_{-0.09}$  \\[5pt] 
    A2199    &  LE & 238.9$\pm$4.0   &   0.351       &   -    & 26.05$^{+2.41}_{-3.07}$ &  72.48$^{+5.43}_{-6.82}$ & 558.1  & 2.54$^{+0.12}_{-0.18}$ \\[5pt] 
    IC1262   &  LE & 184.7$\pm$4.8   &   0.211       &   -   & 3.04$^{+0.25}_{-0.24}$   &  12.52$^{+0.28}_{-0.36}$  &  283.0  & 0.33$^{+0.01}_{-0.01}$  \\[5pt]  
\enddata
\tablenotetext{*}{No clear BCG (and cluster appears to be highly out of equilibrium) -- mass estimates accordingly less certain.  Given {\em isopot} value are for closest bright galaxy to cluster center but a minimal isothermal component is used during fitting.}
\tablenotetext{+}{Potentially heightened BCG luminosity/equivalent stellar dispersion due to possible ongoing merger.}
\tablenotetext{1}{denotes a 2MASS drop-out.}
\tablecomments{Note that a Beta model was used to account for cluster emission outside of the outermost annulus in instances where there was still clearly cluster X-ray emission beyond this.  Errors on M$_{2500}$ do not include the additional 5\% systematic uncertainty.  See text for more details.}
\end{deluxetable*}


\section{Notes on Mass Profiles of Individual Clusters} \label{Appendix:Notes_on_Individual_Clusters}

Cluster mass profiles were calculated according to the prescription outlined in Section \ref{Section:Mass_Fitting}.  Here we give additional notes on a subset of systems where either special attention was required or we found substantial differences from previously reported masses. \\
 \\
{\bf A2626:} A relatively small cool-core cluster.  \citet[][]{Zhao13} report an M$_{\rm 500} = 1.81 \pm 0.14\times10^{14}M_{\odot}$ at R$_{\rm 500}$=850~kpc, which is 25\% higher than our mass at an equivalent radius, though we note that extrapolation of our profiles beyond R$_{\rm 2500}$ is uncertain. \\
{\bf A3667:}  This is a non cool-core cluster that is tagged as a merger in \citet[][]{Vikhlinin09}, who found M$_{\rm 500} = 6.74 \pm 0.09\times10^{14}M_{\odot}$.  The total cluster mass reported could be underestimated as a result of this system being substantially out of hydrostatic equilibrium. \\
{\bf A2142:}  A seemingly relaxed non cool-core cluster that contains a distinct cold front \cite[][]{Owers11}.  Nevertheless, our calculated mass is in reasonable agreement with the M$_{\rm 500} = 11.70 \pm 0.45\times10^{14}M_{\odot}$ reported by \citet[][]{Vikhlinin09} \\
{\bf A3376:} This non cool-core cluster does not have a clear central BCG and was previously found to not be described well by either an NFW or King model \cite[][]{Ettori02}.  A convergent fit is found when the isothermal component is minimised (consistent with negligible central stellar component) though the large NFW scale radius means that the fit is essentially reverting to a power-law, suggesting that this object is perhaps a small group.  The data is insufficient to recover central ICM properties inwards of $\sim$15~kpc and so the uncertain inner mass profile is not considered overly concerning.  Our reported M$_{\rm 2500}$ is in reasonable agreement with the ACCEPT mass profile for this object \cite[][]{Cavagnolo09}. \\
{\bf A1644:} This system is a complex merging cool-core \cite[][]{Reiprich04}, containing a major substructure approximately 700~kpc north-east of the main cluster that itself contains a spiral surface brightness feature indicative of ongoing sloshing.  The BCG is a very large cD extending $\sim$80~kpc.  Our modelling is limited to an outermost radius of 424~kpc, beyond which contamination from the substructure causes unstable fits, though a $\beta$ component is still not favored.  Our recovered M$_{\rm 2500}$ appears low compared to the total cluster masses (M$_{\rm 500}$) of $4.0$--$4.5\times10^{14}M_{\odot}$ reported by \citet[][]{Vikhlinin09} (X-ray) and \citet[][]{Girardi98} (optical), though is in reasonable agreement with the mass at same radius reported in ACCEPT \cite[][]{Cavagnolo09}.   Note that whilst our mass estimate is for the main cluster structure only and therefore likely to underestimate the total mass if extrapolated beyond R$_{\rm 2500}$, the mass is relevant for the dynamical times required within our region of interest.  This is particularly true at low altitudes where the large BCG dominates the potential. \\
{\bf A2319:} A hot and rather massive non cool-core cluster.  Extrapolation of the profile is in agreement with the M$_{\rm 500}$ and M$_{\rm 200}$ masses found by \citet[][]{Reiprich02}. \\
{\bf A1413:} We recover a higher M$_{\rm 2500}$ than \citet[][]{Allen08} (M$_{\rm 2500}\sim3.5\times10^{14}M_{\odot}$ at 599~kpc) and \citet[][]{Vikhlinin06} ($\sim3.0\times10^{14}M_{\odot}$).  The BCG of this cluster appears to be undergoing a merger and hence its inferred isothermal velocity dispersion may be biased high.  To test for this we re-fit but with the fixed isothermal component halved, and find that the NFW component compensates to return an almost identical M$_{\rm 2500}$. \\
{\bf A2034:} This is a large diffuse cluster without an obvious central BCG.  The best fit NFW is recovered when minimising the isothermal component, consistent with the lack of a central stellar potential. \\
{\bf PKS0745-191:} Cool-core cluster with clear cavity system.  \citet[][]{Main17} and \citet[][]{Allen08} report marginal disagreement in mass, reporting M$_{\rm 2500} = 3.3\times10^{14}M_{\odot}$ and M$_{\rm 2500} = 4.8\times10^{14}M_{\odot}$ at R$_{\rm 2500} = 600, 680~kpc$ respectively.  Our recovered cluster mass lies between these values. \\
{\bf Hercules A:} The BCG of this cluster contains the powerful FR-I radio source 3C348.  Our M$_{\rm 2500}$ is approximately half that reported by \citet[][]{Main17} ($\sim2.1\times10^{14}M_{\odot}$ at R$_{\rm 2500} \approx 500~kpc$) though is in good agreement with the M$_{\rm 2500} = 1.23\times10^{14}M_{\odot}$ at 423~kpc reported by \citet[][]{Comis11}. \\
{\bf A2204:}  A well studied and massive cool-core cluster.  Two M$_{\rm 2500}$ values could be found in the literature, with $4.1\times10^{14}M_{\odot}$ and $5.5\times10^{14}M_{\odot}$ reported by \citet[][]{Allen08} and \citet[][]{Main17} at R$_{\rm 2500} \approx 630$ and $700~kpc$ respectively.  Further, \citet[][]{Vikhlinin09} reported M$_{\rm 500} \approx 8.9\times10^{14}M_{\odot}$.  The inferred velocity dispersion appears high, and is perhaps biased by the presence of strong, extended, and ongoing star formation in the BCG \citet[][]{Oonk11}.  Our M$_{\rm 2500}$ is in agreement with the \citet[][]{Allen08} number, lying between this and the \citet[][]{Main17} value. \\
{\bf A1991:} Small cool-core cluster with a relatively large BCG.  \citet[][]{Vikhlinin06} reported M$_{\rm 2500} \approx 0.63\times10^{14}M_{\odot}$ and M$_{\rm 500} \approx 1.23\times10^{14}M_{\odot}$, with \citet[][]{Comis11} finding M$_{\rm 2500} \approx 0.32\times10^{14}M_{\odot}$ at R$_{\rm 2500}$ = 279~kpc.  Our outermost annulus extends to 779.1~kpc although there appears to be emission beyond this, and indeed the inclusion of a $\beta$ parameter significantly improves the fit.  This increases our M$_{\rm 2500}$ from $\sim$0.43-- to $\sim0.55\times10^{14}M_{\odot}$, in agreement with the \citet[][]{Vikhlinin06} value. \\
{\bf Zw7160:} This cluster is also known by the names MS 1455.0+2232 and ZwCl 1454.8+2233.  \citet[][]{Comis11} find an M$_{\rm 2500} \approx 1.53\times10^{14}M_{\odot}$ at 436~kpc.  \citet[][]{Martino14} used both {\it XMM-Newton} and {\it Chandra} data to independently calculate two estimates of M$_{\rm 2500}$ for this cluster, finding $M_{\rm 2500} \approx 1.58 \times10^{14} M_{\odot}$ and $M_{\rm 2500} \approx 1.89 \times10^{14} M_{\odot}$ respectively.  The relatively high redshift of this cluster means its ICM can be traced to a radial distance of almost 3~Mpc.  Modelling out to R$\approx$2742~kpc gives an M$_{\rm 2500}$ in approximate agreement with the previously reported values.  However, our fits are poor since there is no discernible cluster signal beyond $\sim$1.2~Mpc. Restricting ourselves to R$\lesssim$1.2~Mpc we recover a statistically improved fit, though our mass is now slightly above previously found values.  We note that our restricted radial range is reasonably close to that used by \citet[][]{Martino14}. \\
{\bf A1758:} This is a complex, distorted, and diffuse non cool-core cluster with no obvious center or BCG.  Further complicating the system is a secondary cluster about 2~Mpc to the South.  If we truncate our fitted region to $\lesssim$875~kpc so as to exclude the secondary object, include a $\beta$ parameter to allow for excluded emission, and minimise the isothermal component in recognition of the lack of a clear stellar component at the cluster center then we recover a convergent fit.  Our M$_{\rm 2500}$ is higher than that found by \citet[][]{Comis11} (M$_{\rm 2500} = 0.052\times10^{14}M_{\odot}$ at 144~kpc). \\

\bibliography{refs}

\end{document}